%% file: Paper.tex
\documentclass[9pt]{report}
\usepackage{csvsimple}
\usepackage{abstract}

\usepackage{pdfpages}
\usepackage{color,soul}
\usepackage{lipsum}  
\usepackage{xcolor}
\usepackage{arydshln}
\usepackage{diagbox}
\usepackage{afterpage}
\usepackage{amssymb} 
\usepackage{epigraph}
\usepackage{siunitx} 
\usepackage[english]{babel}
\usepackage{graphicx}        
\usepackage[numbered,autolinebreaks,useliterate]{mcode}
\usepackage[center]{caption}
\usepackage[font={small,it}]{caption}

\usepackage{subcaption}
\usepackage{csvsimple}
\usepackage{geometry} 
\usepackage{textcomp}
\usepackage{csquotes}
\usepackage{epsfig}
\usepackage{url}
\usepackage{floatrow}
\usepackage{enumitem}
\usepackage{appendix}
\usepackage{marginnote}
\usepackage{amsmath}
\usepackage{siunitx}
\usepackage[makeroom]{cancel} 
\usepackage{pgfplots}
\pgfplotsset{width=7cm,compat=1.8}
\usepackage{pgf,tikz,tikz-3dplot}
\usetikzlibrary{spy}
\usetikzlibrary{backgrounds}
\usetikzlibrary{decorations}
\usetikzlibrary{math}
\usetikzlibrary{intersections} 
\usepackage{booktabs}

\usepackage{pdfpages}

\usepackage[style=numeric-comp,sorting=none,maxbibnames=10]{biblatex}
\DeclareNameAlias{author}{last-first}
\DeclareFieldFormat{labelalphawidth}{#1}
\usepackage{helvet}

\usepackage{varioref}
\usepackage[hidelinks]{hyperref}
\hypersetup{
	pdfcreator = {},
	pdfproducer = {}
}
\usepackage{cleveref}

\usepackage{colortbl} 

\definecolor{Joulegreen}{RGB}{29,143,144}
\definecolor{AUmagenta}{RGB}{95,0,48}
\definecolor{AUorange}{RGB}{238,127,0}
\definecolor{AUpink}{RGB}{226,0,122}
\definecolor{AUpink1}{RGB}{226,73,155}
\definecolor{AUpink2}{RGB}{226,93,164}
\definecolor{AUpink3}{RGB}{226,111,172}
\definecolor{AUpink4}{RGB}{226,128,180}
\definecolor{AUgreen}{RGB}{139,173,63}
\definecolor{AUdarkgreen}{RGB}{119,153,43}
\definecolor{AUblue}{RGB}{0,36,70}
\definecolor{AUbluegrey}{RGB}{127,161,197}
\definecolor{AUlightblue}{RGB}{55,159,203}
\definecolor{mygrey}{RGB}{140,140,140}
\definecolor{mygrey2}{RGB}{100,100,100}
\definecolor{mypink}{RGB}{241,156,187}
\definecolor{myred}{RGB}{127,0,0}
\definecolor{myyellow}{RGB}{255,220,0}
\definecolor{mygreen}{RGB}{28,172,0} 
\definecolor{mylilas}{RGB}{170,55,241}

\usepackage{titlesec}
\usepackage{xcolor}
\usepackage{appendix}
\usepackage{titletoc}
\usepackage{changepage}
\usepackage{fancyhdr} 

\usepackage[T1]{fontenc}
\usepackage{titlesec, blindtext, color}
\titleformat{\chapter}[block]{\large}{\textcolor{white}{\thechapter}\hsp}{0pt}{\color{white}{\large}} 
\titlespacing*{\chapter}{0pt}{-20pt}{40pt}
\titleformat{\section}
  {\Large \color{black} \sffamily\bfseries}{\thesection}{1em}{}
\titleformat{\subsection}
  {\large \color{black} \sffamily\bfseries}{\thesubsection}{1em}{}
\titleformat{\subsubsection}
  {\normalsize\color{black} \sffamily\bfseries\slshape}{\thesubsubsection}{1em}{}
  
\addto\captionsenglish{
  \renewcommand{\contentsname}%
    {Table of Contents}%
}
\addto\captionsenglish{
  \renewcommand{\listfigurename}%
    {List of Figures}%
}
\addto\captionsenglish{
  \renewcommand{\listtablename}%
    {List of Tables}%
}

\addto\captionsenglish{} 
\addto\captionsenglish{} 

\usepackage{amssymb}
\usepackage[nocfg]{nomencl}
\makenomenclature
\usepackage{etoolbox}

\usepackage{lastpage} 
\newcommand{\whiteline}[1]{%
	
	\parbox{\textwidth}{
		\centering
		\parbox{0.6\linewidth}{
			\centering
			\color{white}\rule{\linewidth}{1.5pt}
		}
	}
}

\usepackage{adjustbox}

\usepackage{multicol}

\newcommand{\beginsupplement}{%
	\setcounter{table}{0}
	\renewcommand{\thetable}{S\arabic{table}}%
	\setcounter{figure}{0}
	\renewcommand{\thefigure}{S\arabic{figure}}%
}

\begin{document}

\include{./Chapters/text}
\include{./Chapters/supplementary}















\end{document}

%% file: Chapters/text.tex
\chapter*{Article}
\vspace{-1 cm} 
{\huge Future operation of hydropower in Europe under high renewable penetration and \newline climate change \par}



\vspace{1\baselineskip}

Ebbe Kyhl Gøtske$^{a}$, Marta Victoria$^{a,b}$

\vspace{1\baselineskip}

{\small $^{a}$ Department of Mechanical and Production Engineering, Aarhus University \\
$^{b}$ iCLIMATE Interdisciplinary Centre for Climate Change, Aarhus University}
\vspace{1\baselineskip}




\subsection*{SUMMARY} 
{\textcolor{black}{\fontfamily{qcr} \hspace{-0.2cm} 
		The balancing provided by hydropower reservoirs is essential in the transition towards a decarbonised European energy system, but the resource might be impacted by future climate change. In this work, we first analyse the hydropower operation needed to balance a wind and solar dominated European energy system, to signify whether and to what extent hydropower is required to operate differently due to the decarbonisation of the energy system. Second, we apply runoff data achieved with 10 dynamically downscaled climate models with $0.11^\circ \times 0.11^\circ$ horizontal and daily resolution to project the future reservoir inflow at three CO$_2$ emissions scenarios: low (RCP2.6), mid (RCP4.5), and high emissions (RCP8.5). We show that the decarbonised energy system increases the ramp rates and seasonality of the hydropower operation. Despite large interannual and intermodel variability, we found a significant change in annual inflow due to climate change in 20 out of 22 European countries at the mid and high emissions scenarios. The seasonal profile, as well as the frequency and duration of droughts and floods, is also projected to be impacted.}} 

\subsection*{INTRODUCTION}


The power sector is facing a major transformation from using fossil fuel generators to renewable and carbon-neutral ones. The European Green Deal proposed by the European Commission \parencite{EUGreendeal2019} envisions a climate neutral Europe by 2050. This target could be achieved by relying on wind and solar power generation balanced by hydropower and other dispatchable generators \parencite{Victoria2020}. Hydropower is a substantial part of the European electricity generation (accounting for 16\% in 2018 \parencite{iea2021}) and will remain an important energy source in the future. Due to the variable nature of wind and solar generation, their increasing penetration in power grids strengthens the need for balancing and demands a modifided operation of hydropower plants \parencite{PFEIFFER2021110885}. Some hydropower installations can, however, benefit from wind and solar since they are able to contribute to a seasonal dispatch similar to the natural river discharge which is favourable for the river ecosystem \parencite{Sterl2021}.\\


The renewable energy supply is furthermore affected by the changing climate conditions \cite{Cronin2018,Yalew2020,EEA2019}. Most European regions are at the end of the century projected to encounter increased wind correlation lengths \parencite{Schlott2018} and higher probability of low wind regimes \parencite{WEBER201822}. For solar power, the radiation is projected to decrease in the Northern Europe \parencite{Schlott2018,Jerez2015}. Schlot et al. \parencite{Schlott2018} investigated the future European power system under climate change impacted series for solar, wind and hydro and found that the cost-optimal contribution of solar increases as a result of climate change. This is in agreement with the results found by Fonseca et al. \parencite{Fonseca2021} for the Southest US. Previous studies show that hydropower is particularly sensitive to climate change \parencite{LEHNER2005,Turner2017,Vanvliet_forecast2016,Vanvliet2013,Schlott2018,Schaefli2007}. Turner et al. \parencite{Turner2017} apply three general circulation models (GCMs) at a low and high emissions scenario in a high-fidelity dam model with a variable hydraulic head to project the potential hydropower production for 1,593 globally distributed hydropower reservoirs at the middle of the century. The authors select hydropower reservoirs whose dam and turbine specifications (e.g. dam height, storage capacity, upstream catchment area, maximum turbine flow rate, etc.) are available. Van Vliet et al. \parencite{Vanvliet_forecast2016} use a higher number of hydropower plants (24,515) and GCMs (5) but less information (fixed hydraulic head) for each hydropower plant. Both studies, which use climate model data of daily temporal and $0.5^\circ \times 0.5^\circ$ ($\approx \SI{50}{km} \times \SI{50}{km}$) horizontal resolution, project a north-south European division in which the Nordic countries gain an increase in annual hydropower production whereas the Mediterranean and Balkan countries suffer from a reduction. However, the two studies do not agree on the outlook for global hydropower. The former projects a change in annual electricity production of $\pm 5\%$ depending on the GCM, whereas the latter projects a robust (across all GCMs) decrease in the range from $0.4\%$ to $6.1\%$. Since projections of hydropower impacts at a global and regional scale are known to vary \parencite{Yalew2020,Cronin2018}, climate model uncertainty has been identified as a significant research gap \parencite{Cronin2018}. Moreover, both aforementioned analyses focus on investigating the impact of climate change on annual figures and assume that the hydropower operation is independent of the rest of the power system.\\


The available energy for electricity production at a hydropower plant is decided by the local characteristics such as topography, snow formation, temperature, and precipitation which might not be adequately represented in the GCM due to its coarse horizontal resolution. By downscaling the climate model simulation results, using a regional climate model (RCM), the local and regional characteristics obtain a higher representativeness \parencite{Christensen}. A comprehensive comparison of the current supply of RCMs for Europe was conducted by Christensen and Kjellström \parencite{Christensen} in which a $5 \times 5$ GCM-RCM matrix was analysed for temperature, wind speed, and precipitation. It showed that the climate model signal was generally more influenced by the choice of GCM than RCM, but that the RCM has higher influence in mountainous regions due to the added surface detail. In this work, we concentrate on runoff since the inflow in reservoirs is highly correlated to the runoff within the upstream basins of the reservoir \parencite{Liu2019}. Furthermore, we use an advanced conversion scheme to model inflow from climate model runoff data in European reservoirs. To our knowledge, a comprehensive analysis of a large climate model ensemble of runoff with an advanced inflow conversion scheme has not been conducted before. Two additional novelties of our study are the use of several RCMs with high resolution and the analysis of climate change impacts on hydro seasonal profiles, as well as frequency and duration of droughts and floods.\\









In this work, we investigate the following research question: How is the operation of reservoir hydropower plants in Europe expected to change in the future? We first analyse balancing requirements due to high wind and solar penetration in the grids, and, second, the changed inflow caused by climate change. The former is based on the 2050 decarbonised cost-optimal European energy system \parencite{Victoria2020}, and the latter on 10 combinations of GCM-RCM runoff climate model data acquired from the Coordinated Regional Climate Downscaling Experiment (CORDEX) \parencite{Giorgi2009}. An emphasis is put on the variability caused by the differences between the climate models (intermodel variability) and within the considered period (interannual variability) to quantify the robustness and significance of the resulting projection. The time series that we produce in this project are released under an open license. They include hourly-resolved time series of inflow for 22 European countries at the beginning of the century (BOC) from 1991 to 2020 and the end of the century (EOC) from 2071 to 2100, for three CO$_2$ emissions scenarios (RCP2.6 RCP4.5 and RCP8.5), for 10 different climate models, as well as the average of the model ensemble.

\subsection*{METHODS}

\subsubsection{Sector-coupled European energy model with high wind and solar penetration}

The historical operation of hydropower in European countries is compared to the cost-optimal operation that will be required as the power system decarbonises. For the latter, hydropower dispatch time series in 2050 obtained with PyPSA-Eur-Sec are used \parencite{Victoria2020}. The PyPSA-Eur-Sec model optimises the capacity and dispatch of every generation, storage and transmission technology subject to a global CO$_2$ constraint. The model assumes perfect competition and foresight as well as long-term market equilibrium. The system including the electricity and heating sectors is optimised every 5 years from 2020 to 2050 following a myopic approach, i.e. in every optimisation step, only information regarding that year is available. Technologies installed in previous time steps remain in the model until they reach the end of their lifetime. The CO$_2$ constraints in every time step correspond to a carbon budget of 21GtCO$_2$ which enables avoiding human-induced warming above 1.75$^\circ$C with a probability of >66\%, assuming current sectoral distribution for Europe, and equity sharing principle among regions. The Baseline - Early and Steady pathway in Victoria et al. \parencite{Victoria2020} is selected.

\subsubsection{Modelling hydropower inflow time series under various climate scenarios}

\begin{figure}[ht]
	\centering
	\includegraphics[width=12cm]{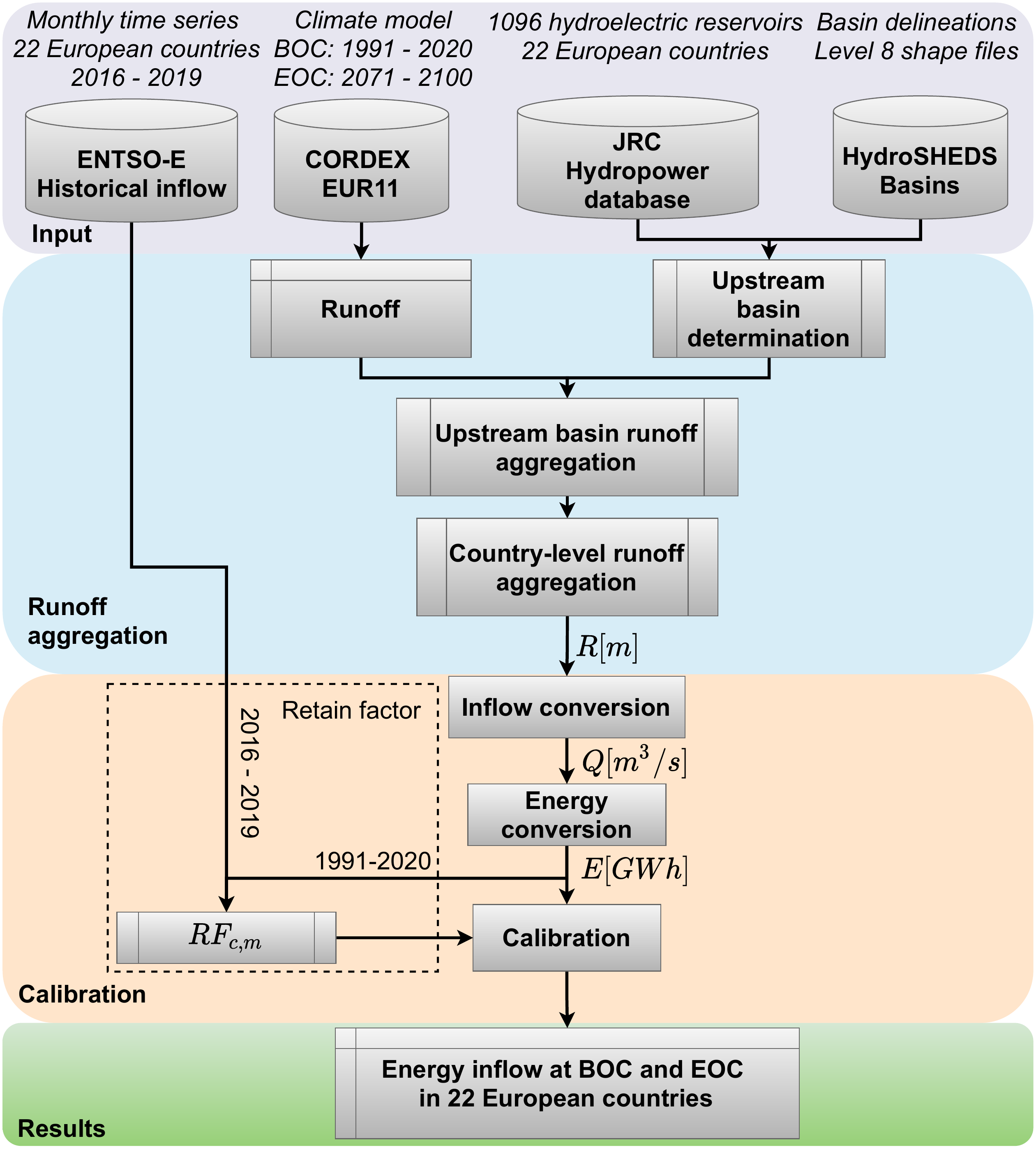}
	\caption{Scheme of the methodology to obtain hourly resolved inflow time series at the two periods, beginning of the century (BOC) from 1991 to 2020 and end of the century (EOC) from 2071 to 2100.}	
	\label{fig:methods}
\end{figure}

Figure \ref{fig:methods} presents the methodology applied in this study to obtain country-aggregated energy inflow time series from climate model data. The time series are obtained using the location of the hydropower plants \parencite{jrc}, the geometry of the upstream basins \parencite{Lehner2013}, and the open-source package Atlite \parencite{ATLITE}. A detailed description is provided in Note \hyperref[s:methods]{S1}. The methodology to obtain inflow time series was introduced in Liu et al. \parencite{Liu2019}, where it was used to obtain historical time series in China using reanalysis weather data.

Climate model data is acquired from EURO-CORDEX \parencite{Cordex_database} driven by general circulations models from the Coupled Model Intercomparison Project - Phase 5 (CMIP5) with $0.11^\circ \times 0.11^\circ$ ($\approx \SI{12}{km} \times \SI{12}{km}$) horizontal resolution. Every climate model is given by a combination of a GCM and an RCM. An ensemble consisting of five different GCMs and two RCMs is applied in this study. Furthermore, each combination is investigated at three different representative concentration pathways (RCPs), RCP2.6, RCP4.5, and RCP8.5 \parencite{ipcc2014}. The 5x2x3 matrix of combinations is shown in Table \ref{tab:gcm_rcm_matrix}. Although some climate models do not contain data for all RCPs, the three scenarios are well represented with a total of 6 models assuming RCP2.6, 8 RCP4.5, and 10 RCP8.5, which constitutes an ensemble of 24 climate models. As a preliminary step, the interannual, inter-RCM, inter-GCM, and climate change-induced variability of the runoff is evaluated for the RCP8.5 scenario. 
As a frame of reference, we define the period from 1991 to 2020 as beginning of the century (BOC) and from 2071 to 2100 as end of the century (EOC). The climate model runoff is indexed with the notation $\bar{R}_{ijkl}$ in which index $i$ represents the considered period ($i=1$ for BOC and $i=2$ for EOC), $j$ the GCM, $k$ the RCM, and $l$ the RCP. The bar indicates the mean annual value within a period of 30 years.\\

\begin{table}[ht]
	\centering
	\caption{Combination of general circulation models (GCMs), regional climate models (RCMs), and representative concentration pathways (RCPs) considered in this study.}
	\label{tab:gcm_rcm_matrix}
	\resizebox{\textwidth}{!}{
	\begin{tabular}{@{}llccccc@{}}
		\toprule
		\backslashbox{RCM}{GCM} & & MPI-ESM-LR\parencite{Giorgetta2013} & EC-EARTH\parencite{Hazeleger2012} & CNRM-CM5\parencite{Voldoire2013} & HadGEM2-ES\parencite{Collins2011} & NorESM1-M\parencite{Bentsen2013} \\ \midrule
		RCA4\parencite{Samuelssson2011} & RCP2.6 & x & x & & x & x \\
		& RCP4.5 & x & x & x & x & x \\
		& RCP8.5 & x & x & x & x & x \\
		HIRHAM5\parencite{Christensen2007} & RCP2.6 & & x & & x & \\
		& RCP4.5 & & x & & x & x \\
		& RCP8.5 & x & x & x & x & x \\
		\bottomrule
	\end{tabular}
	}
\end{table}


The climate change effect $S$ is determined by comparing the runoff from EOC with BOC. E.g. the absolute change in mean daily runoff when assuming RCP8.5 is calculated as:
\begin{equation}\label{eq:climate_change_runoff}
	S^R_{jk} = \bar{R}_{2jk3} - \bar{R}_{1jk3}
\end{equation}

The climate change effect on the mean annual inflow is evaluated by calculating the relative change:
\begin{equation}\label{eq:climate_change_inflow}
	S^E_{jk} = \frac{\bar{E}_{2jk3} - \bar{E}_{1jk3}}{\bar{E}_{1jk3}} \times 100
\end{equation}

\begin{figure}[h!]
	\centering
	\includegraphics[width=\textwidth]{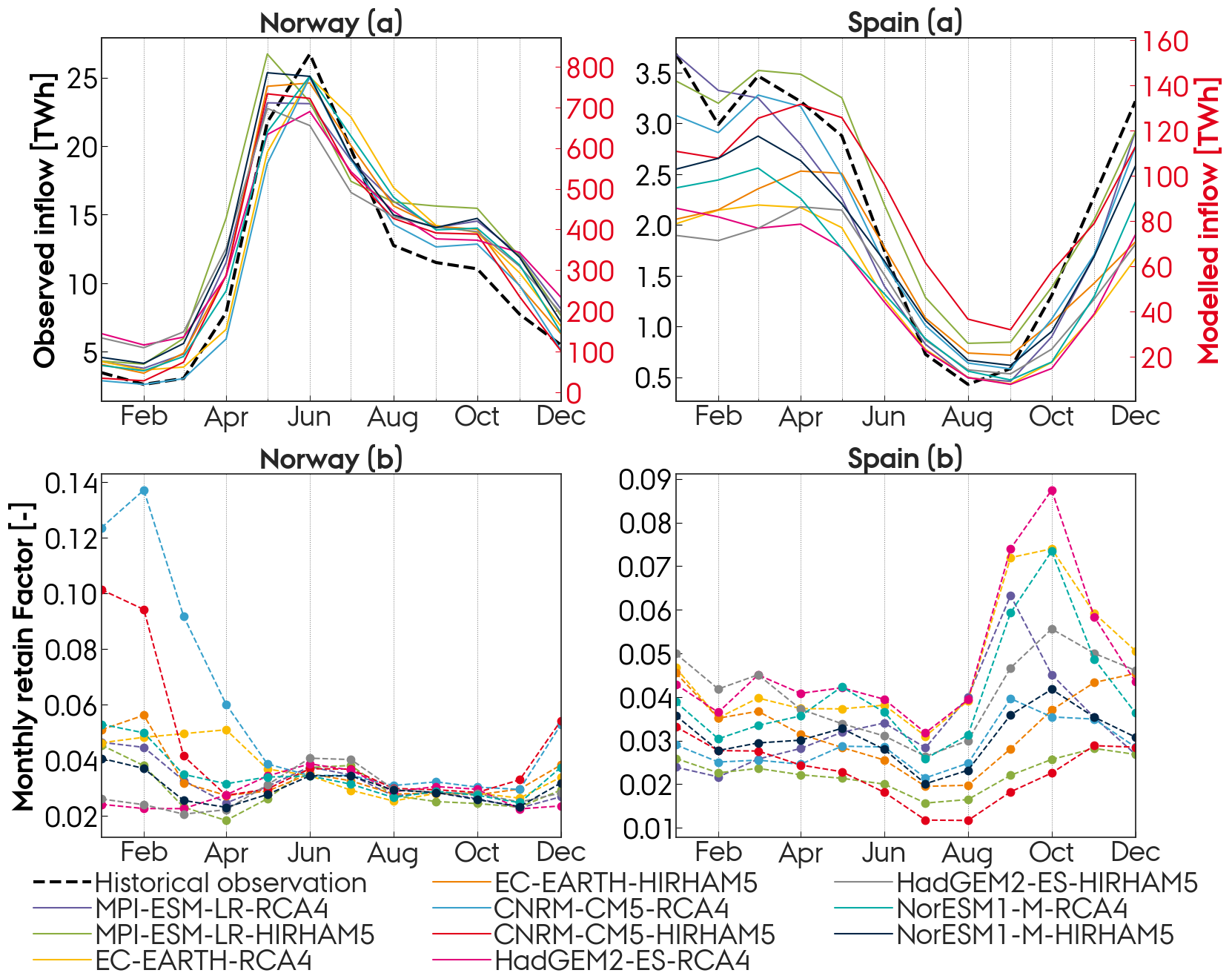}
	\caption{(a) Comparison of modelled, prior to calibration, average monthly inflow at BOC period (1991-2020) and observed inflow (1991-2019), and (b) the corresponding retain factors estimated according to Eq. \ref{eq:month_country_dep_RF} for Norway and Spain. In (a) right axis is modelled and left axis is observed inflow. Similar plots for other countries are shown in Figures \ref{fig:model_evaluation_22} and \ref{fig:retain_factors_22}.}
	\label{fig:model_evaluation}
\end{figure}

The seasonal inflow profiles obtained with the climate models at the BOC period (1991-2020), prior to calibration, are compared with observed inflow (1991-2019) for Norway and Spain in Figure \ref{fig:model_evaluation} which also shows the corresponding monthly retain factors, $RF_{c,m}$. The latter represents the runoff water that does not reach the dam due to evaporation, groundwater infiltration, etc., and it also includes the model bias correction. Similar plots for the remaining countries are shown in Figure \ref{fig:model_evaluation_22} and \ref{fig:retain_factors_22}. The comparison shows a strong correlation (Pearson correlation $r>0.75$) between the modelled and observed mean monthly inflow for 14 countries. Countries in the Balkan Peninsula and north of it show moderate correlations ($r<0.75$).


\begin{figure}[h!]
	\centering
	\includegraphics[width=\textwidth]{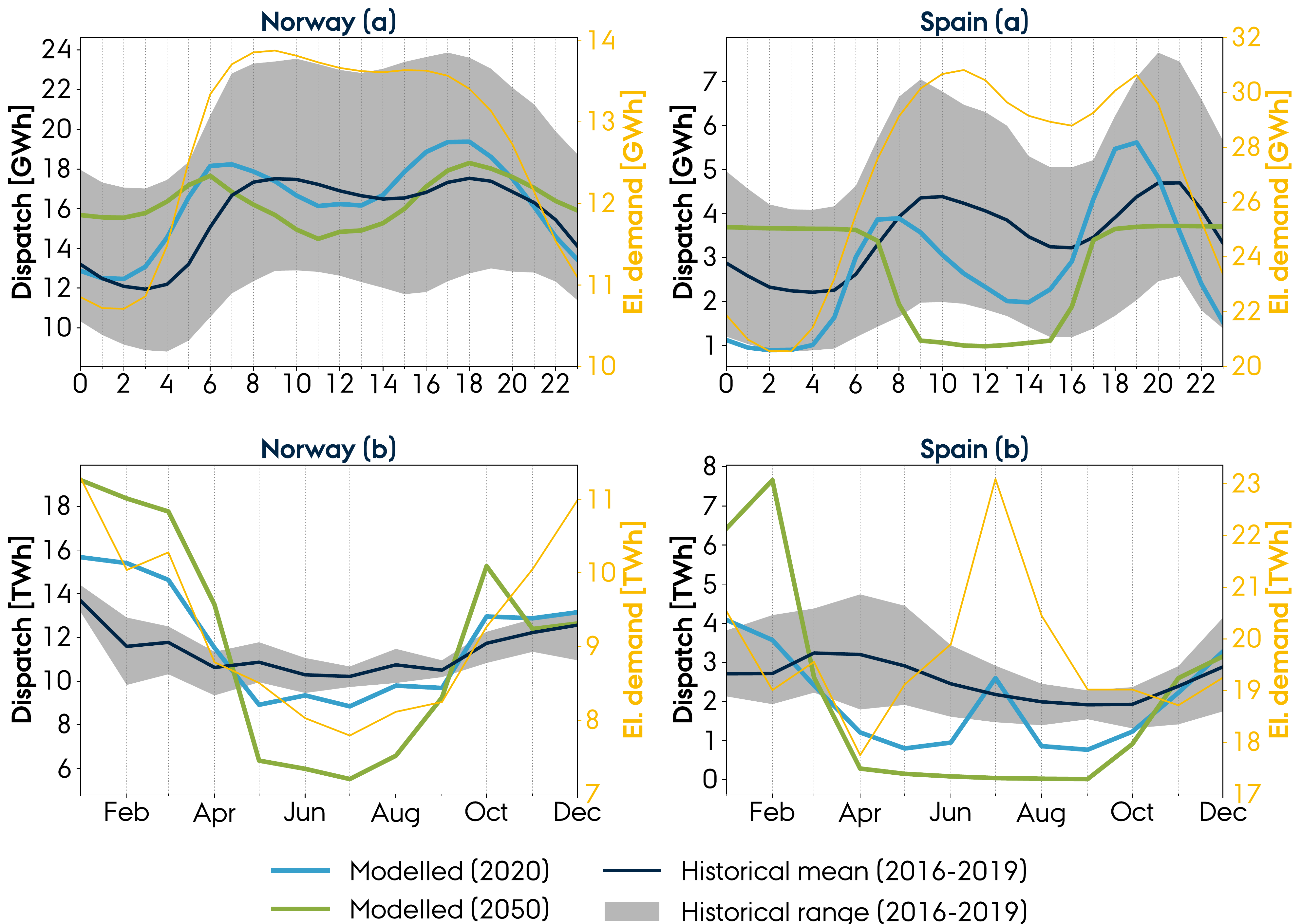}
	\caption{Modelled and observed (a) intraday and (b) seasonal operation of hydropower plants. The shaded grey area indicates the range of the historical observations. Similar plots for other countries are shown in Figure \ref{fig:intraday_season_9}.}
	\label{fig:intraday_dispatch}
\end{figure}




\subsection*{RESULTS AND DISCUSSION}

\subsubsection{Changes in the hydropower operation as wind and solar penetration increases}

For the sake of clarity, we focus the discussion on hydropower in Norway and Spain as countries representative of Northern Europe and Mediterranean climates and which are projected to experience contrasting impacts of the decarbonisation of the energy system \parencite{Victoria2020} and of climate change on hydro resources \parencite{Vanvliet_forecast2016,Turner2017}. Figures in Supplementary material \hyperref[sec:sup_res]{S3} extend the analysis to other countries. Figure \ref{fig:intraday_dispatch} presents the intraday and seasonal profile of the hydroelectricity from historical observations and a cost-optimised model of the European energy system in 2020 and 2050. The modelled intraday operation in 2020 is within the historical range in both countries and captures the morning and afternoon peaks caused by the load curve. The modelled intraday generation in 2050 shows a minimal change in the profile for Norway, whereas Spain shifts from matching the load curve in 2020 to showing a binary operation. The day-night switch can be explained by the large installation of solar capacity in Spain which provides high generation at day but requires balancing at night.\\

	
The modelled hydropower does not resemble the historical seasonal dispatch as well as the intraday profiles. In 2020, modelled dispatch in both countries show a lower magnitude during summer and higher during winter, relative to the historical values. In addition, Spain shows a short generation peak in July caused by high electricity demand. The model assumes perfect foresight (i.e. the weather and inflow are known for the entire year) which might cause too ideal operation of hydropower. By 2050, the high wind and solar penetration forces the energy system to rely more heavily on hydropower during winter. The Spanish hydropower is particularly influenced by the solar penetration, showing almost zero dispatch from April to September, and a short period during winter with a high level of dispatch.\\ 

\begin{figure}[h!]
	\centering
	\hspace{-0.5cm}
	\includegraphics[width=\textwidth]{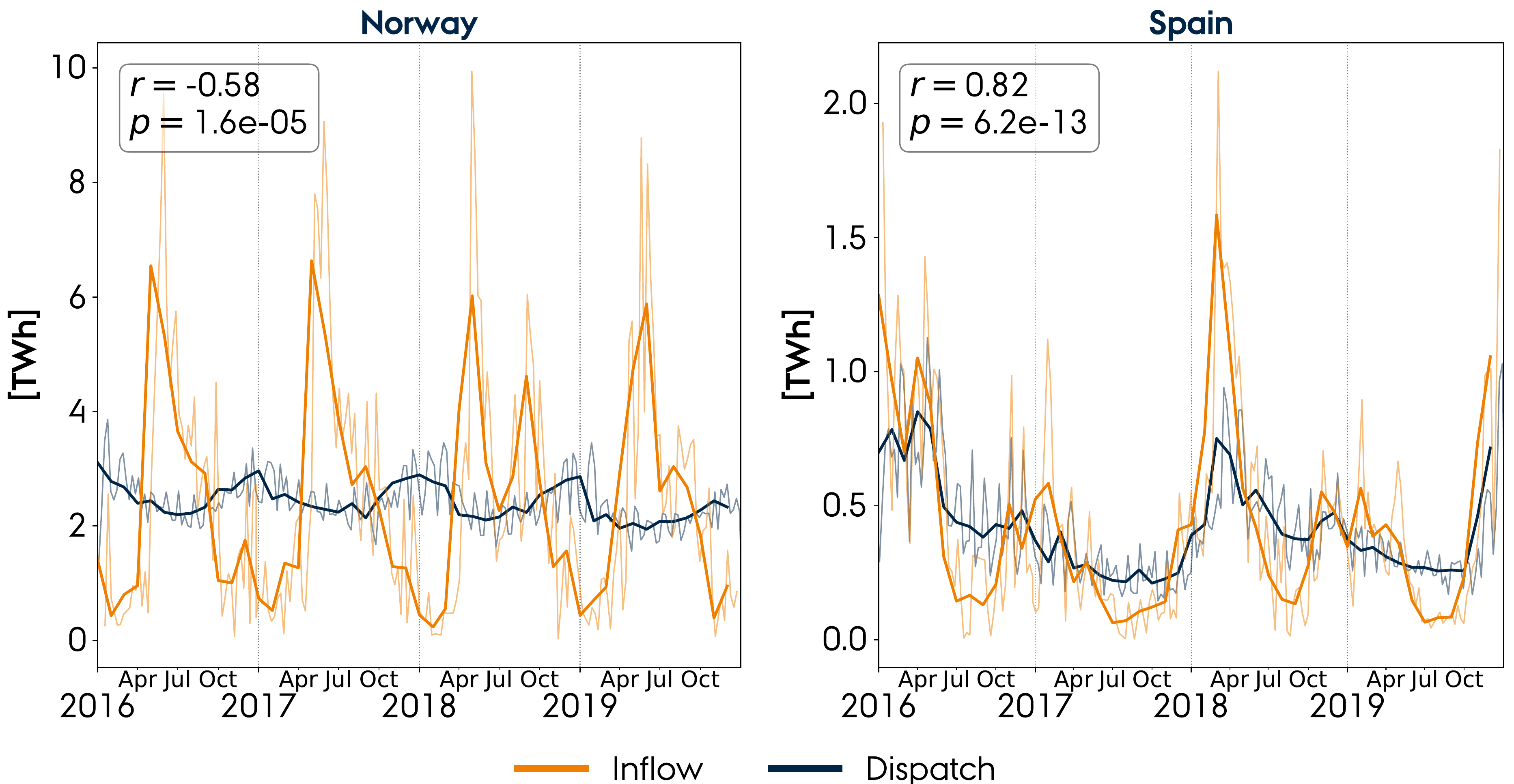}
	\caption{Observed hydropower reservoir inflow and dispatch from 2016 to 2019 from ENTSO-E with $r$ indicating the inflow-dispatch correlation and $p$ the statistical significance. Thin lines indicate weekly values and thick monthly averaged. Similar plots for other countries are shown in Figure \ref{fig:d_i_cor}.}
	\label{fig:d_i_cor_reservoir_ror}
\end{figure}

Figure \ref{fig:d_i_cor_reservoir_ror} illustrates the historical reservoir water inflow and hydropower dispatch, which are negatively correlated in Norway. The inflow in Norway has a summer peak in June caused by the melting of snow, whereas the dispatch peaks in winter; thus, a high share of the hydro resources are saved five months. Conversely, Spain does not show such a capability of shifting the seasonal dispatch based on the historically strong inflow-dispatch correlation. The inflow in Spain, influenced by rain patterns but not by deicing, peaks in winter, while the electricity demand is higher in summer. The strong inflow-dispatch correlation in Spain can thus not be explained by the electricity demand curve, but instead by the lack of seasonal storage capacity. This is also illustrated in Figure \ref{fig:dischargetime_pearson_cor} in which the inflow-dispatch correlation for nine countries shows a strong negative correlation with the country-average size of the reservoirs (ratio between the energy capacity and power capacity of hydropower reservoirs). The increased seasonality in the hydropower operation suggested by the model seems, from a historical perspective, feasible in both Norway and Spain for different reasons. Norway can meet the seasonally shifted hydropower operation due to large reservoirs, whereas Spain is able to comply with increased hydropower generation in winter since it is concurrent with the peak inflow. Figure \ref{fig:intraday_season_9} shows the intraday and seasonal operation for the remaining countries. For Austria and Romania, a similarly increased seasonality is shown in the hydropower operation required by the model. However, the two countries are subject to a peak inflow in summer, and since the hydropower generation in those countries is correlated with the inflow, the modelled operation in 2050 is questionable.
 
The required change in operation may contribute to higher ramp rates which can lead to a reduced lifetime of the hydropower plants due to higher mechanical stresses \parencite{Bakken2002}. See Supplementary Figure \ref{fig:ramp_rates_9}. This is consistent with previous studies \parencite{PFEIFFER2021110885} and \parencite{Gallego2021} which
showed the same effect when only increasing the wind power capacity in a region in the USA. Future operation of hydropower also needs to consider that the artificial flow fluctuations downstream of reservoirs caused by hydropeaking might entail negative consequences for aquatic organisms \parencite{schmutz2018}. 

\subsubsection{Changes in hydropower inflow caused by climate change}

As a prelude to this section, it is noteworthy that we are already witnessing climate change impact on hydropower resources. Figure \ref{fig:historical_cc_no} presents simulated Norwegian reservoir inflow, based on historical precipitation and temperature measurements, from 1958 to 2017, produced by Holmqvist \parencite{Holmqvist2017}, and is a representation of the historical climatic evolvement (energy capacity is assumed constant in the simulation). A clear increasing trend in temperature and annual inflow can be observed.\\ 

\begin{figure}
	\floatbox[{\capbeside\thisfloatsetup{capbesideposition={right,top},capbesidewidth=4cm}}]{figure}[\FBwidth]
	{\caption{Change in mean daily runoff [mm/day] at the EOC period relative to the BOC for the RCP8.5 scenario.}\label{fig:runoff_matrix}}
	{\includegraphics[width=10cm]{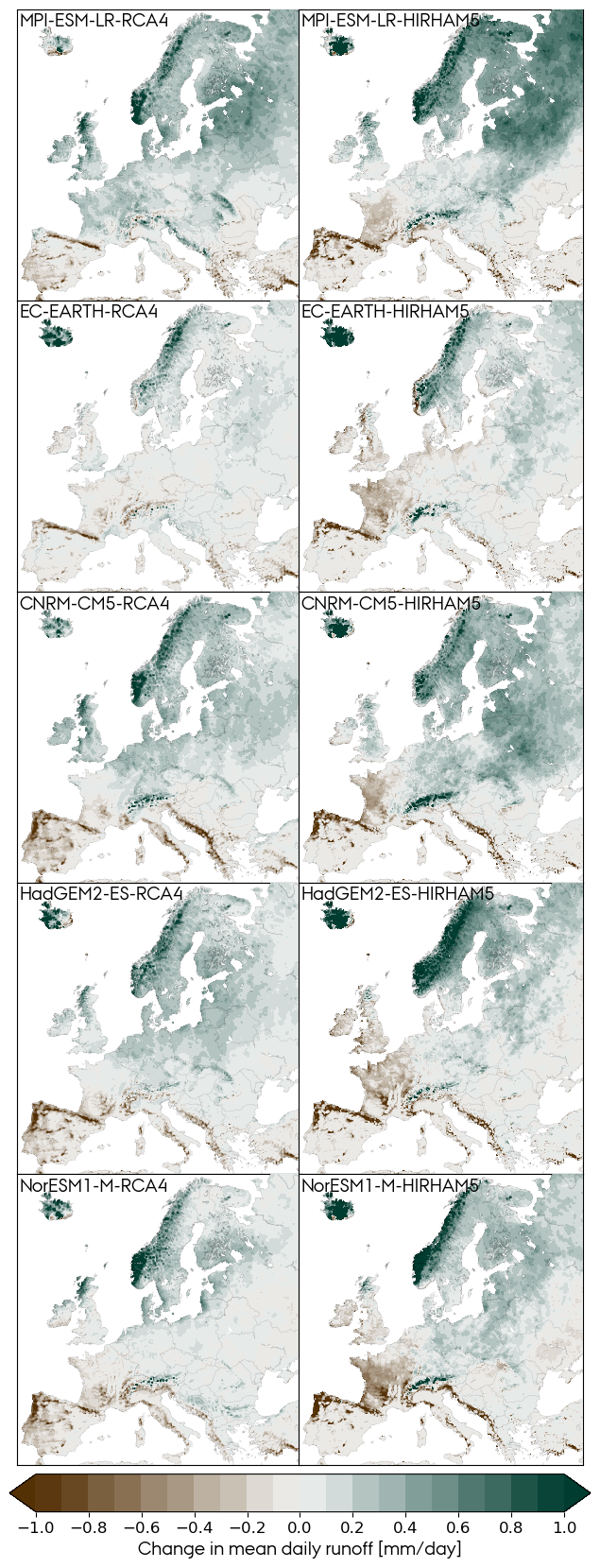}}
\end{figure}
\begin{figure}
	\floatbox[{\capbeside\thisfloatsetup{capbesideposition={right,top},capbesidewidth=4cm}}]{figure}[\FBwidth]
	{\caption{Relative changes in annual country- aggregated inflow at the EOC period relative to the BOC for the RCP8.5 scenario.}\label{fig:inflow_matrix}}
	{\includegraphics[width=10cm]{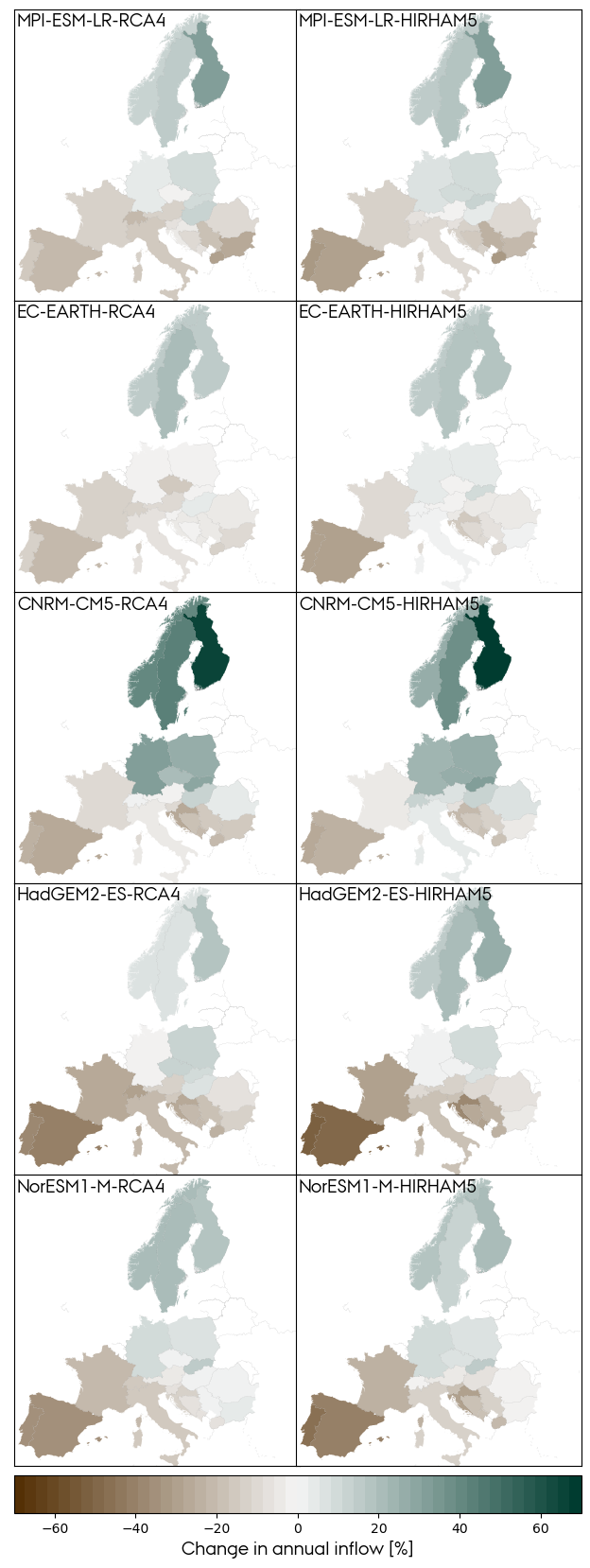}}
\end{figure}

As a preliminary step, we evaluate the absolute change in mean daily runoff at the EOC period relative to the BOC (Eq. \ref{eq:climate_change_runoff}). Figure \ref{fig:runoff_matrix} shows the results for the combinations of 5 GCMs and 2 RCMs for the RCP8.5 scenario. The signal is consistent for the outer areas of the domain, i.e. runoff increases in northern Europe and decreases in the south. The models vary in the magnitude and direction of change in some regions. France is a good example of the latter: Runoff increases or shows negligible changes in combinations using RCA4 as RCM but decreases when HIRHAM5 is used. The corresponding matrix of relative change in annual inflow (Eq. \ref{eq:climate_change_inflow}) derived from the climate model runoff is presented in Figure \ref{fig:inflow_matrix}. Consistently across models, for the RCP8.5 scenario, Mediterranean countries show a significant reduction at the EOC period while Nordic countries show a significant increase. When comparing the runoff and inflow matrix, some considerations must be taken. First, the inflow matrix is based on relative changes, which explains why e.g. Finland in some models shows the largest increase in inflow when it is not reflected on the runoff. Second, the modelled inflow is based on the location of the hydropower plants, see Figure \ref{fig:loc_power_plants}, and e.g. France has most of its plants located in the south, so the inflow is barely affected by the runoff change in the north.\\





\begin{figure}[h!]
	\centering
	\hspace{-1.5 cm}
	\includegraphics[width=1\textwidth]{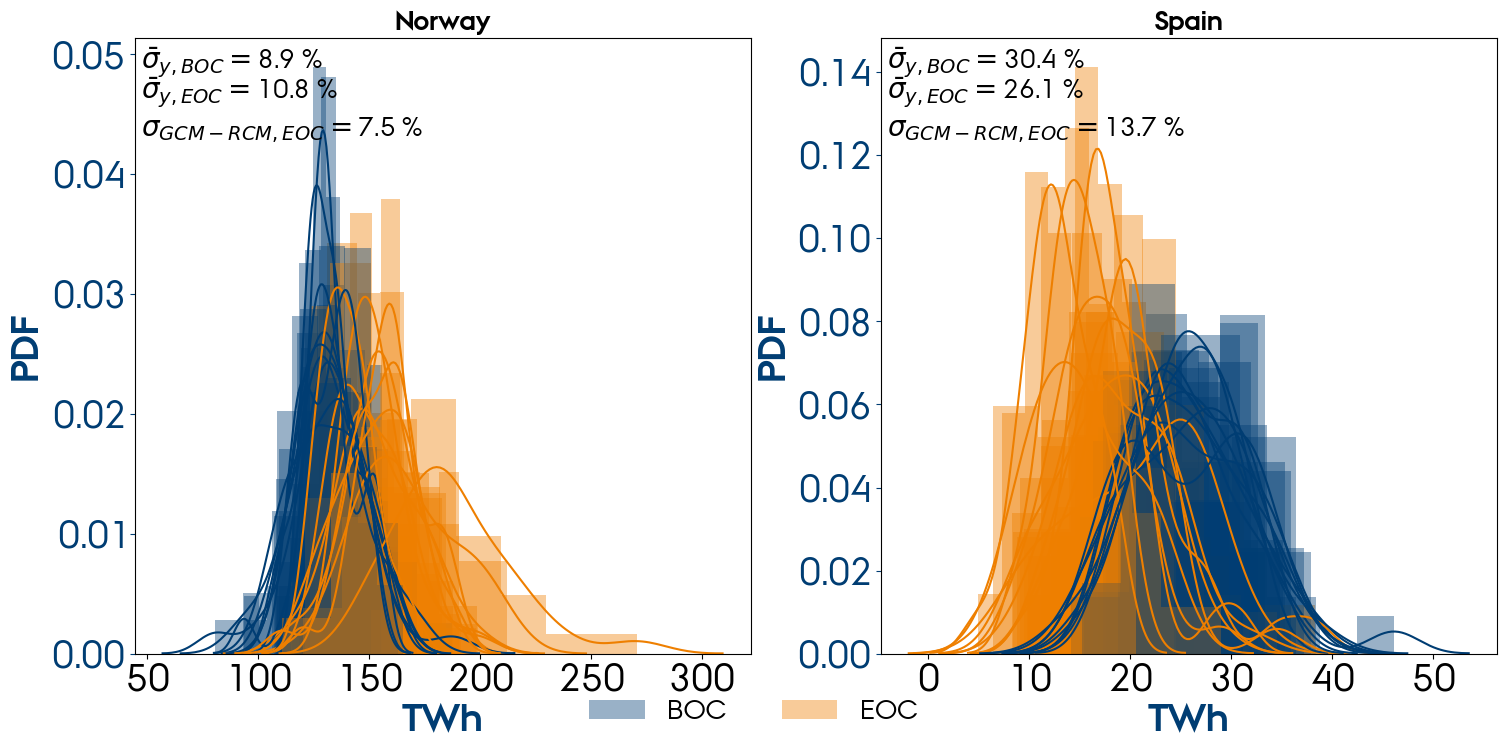}
	\caption{Kernel density estimations of the probability density functions for the annual inflow in Norway and Spain at the BOC (blue) and EOC (yellow) periods for the RCP8.5 scenario with the corresponding interannual variability, $\bar{\sigma}_{y\text{,BOC}}$ and $\bar{\sigma}_{y\text{,EOC}}$, and the intermodel variability at the EOC, $\sigma_{\text{GCM-RCM,EOC}}$.}
	\label{fig:GCM_RCM_dist}
\end{figure}

We now focus on analysing the intermodel and interannual variability. Figure \ref{fig:GCM_RCM_dist} depicts the kernel density estimations of the probability density functions of the annual inflow in Norway and Spain at the BOC and EOC 30-years periods for the 10 different climate models. As Figure \ref{fig:inflow_matrix} anticipates, the climate models all agree on an increase in annual inflow in Norway and a reduction in Spain. The interannual variability, averaged for the 10 climate models, $\bar{\sigma}_y$, is projected to increase from 8.9\% (of the mean annual inflow) at the BOC period to 10.8\% at the EOC in Norway, and decrease from 30.4\% to 26.1\% in Spain. The variability between the 10 climate model (mean) projections, i.e. the intermodel variability ${\sigma}_{\text{GCM-RCM}}$ is 7.5\% for Norway and  13.7\% for Spain at the EOC period (the intermodel variability at the BOC is zero due to calibration with observed inflow). According to analysis of variance (ANOVA) \parencite{Weiss2005}, since the variability between the 10 climate model projections, ${\sigma}_{\text{GCM-RCM}}$, is smaller than the variability within, $\bar{\sigma}_y$, we can treat the 10 climate model projections at the EOC as being from the same population. Based on that result, we group the 10 climate model projections into one ensemble consisting of 30 (years) x 10 (climate models) observations. The variability contributed by changing the GCM or RCM is examined in Supplementary Figure \ref{fig:intermodel}. It is shown that the range of the mean annual inflow $\bar{E}_{EOC}$ and the interannual variability $\bar{\sigma}_{y,EOC}$ is generally larger when varying the GCM than RCM. This result could indicate that the intermodel variability is mainly due to the variation in the GCM, consistent with Christensen and Kjellström \parencite{Christensen}. However, to verify this indication, a larger number of RCMs would be needed.\\


\begin{figure}[h!]
	\centering
	\hspace{-0.5cm} 
	\includegraphics[width=1\textwidth]{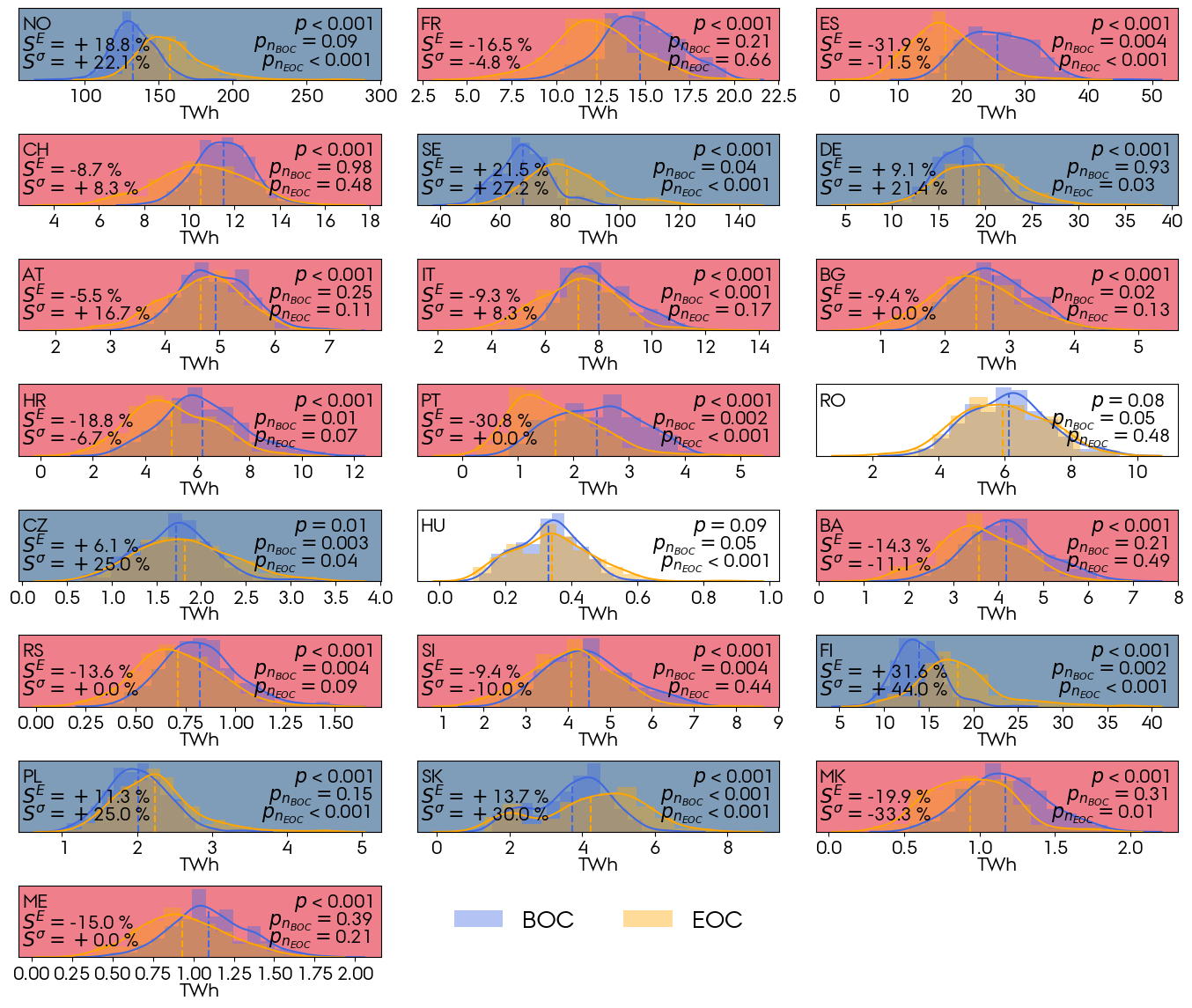}
	\caption{Ensemble distributions of the annual inflow at the BOC and EOC 30-years periods for the RCP8.5 scenario. Dashed lines indicate the mean values of the distributions. The sets are normally distributed if $p_n>0.05$ based on a Shapiro-Wilk test. For the countries with a statistically significant change ($p<0.05$), a blue (red) shade indicates an increase (decrease) in the annual inflow. $S^E$ and $S^\sigma$ correspond to the relative change in annual inflow and interannual variability caused by climate change. For RCP4.5, see Figure \ref{fig:t_test_45}.}
	\label{fig:t_test}
\end{figure}

Figure \ref{fig:t_test} shows the distributions of the annual inflow from the climate model ensemble at the BOC and EOC periods (each consisting of 300 observations) for the 22 countries under analysis and the RCP8.5 scenario. First, a Shapiro-Wilk test \parencite{Shapiro1965} is performed to check if the two sets are normally distributed, which is the case if the probability score $p_{n_{BOC}}$ and $p_{n_{EOC}}$ are above 0.05. France, Switzerland, Austria, Bosnia and Herzegovina, and Montenegro fulfil this condition. Second, a paired t-test is conducted to check if we can reject the null hypothesis stating that the mean difference between the two sets is zero. We apply it to every country, in spite of the indication from the Shapiro-Wilk test that some datasets are not normally distributed, justified by all sets showing similar (bell-shaped) normal distribution characteristics. The null hypothesis of zero difference between the EOC and BOC periods can be rejected for all countries except Romania and Hungary. Figure \ref{fig:t_test} also shows the relative change in annual inflow $S^E$ and mean interannual variability $S^{\sigma_y}$ (normalised with mean annual inflow). The latter is calculated for each model and averaged as in Figure \ref{fig:GCM_RCM_dist}, so $S^\sigma$ does not represent the width of the distributions plotted in Figure \ref{fig:t_test} since they also include the intermodel variability. However, they might be similar since we showed that the interannual variability is larger than the intermodel. Countries in which annual inflow increases also show an enlarged interannual variability, that is, countries that are expected to benefit from more abundant hydropower resources at EOC will also suffer from stronger year-to-year fluctuations. Countries experiencing a reduction in annual inflow do not show the same unanimity regarding the direction of change of the interannual variability: The latter increases for countries in central Europe (Switzerland, Austria, and Italy) but decreases in Mediterranean countries (France, Spain, Croatia, Bosnia and Herzegovina, Slovenia, and North Macedonia).\\

\begin{figure}[h!]
	\centering
	\includegraphics[width=\textwidth]{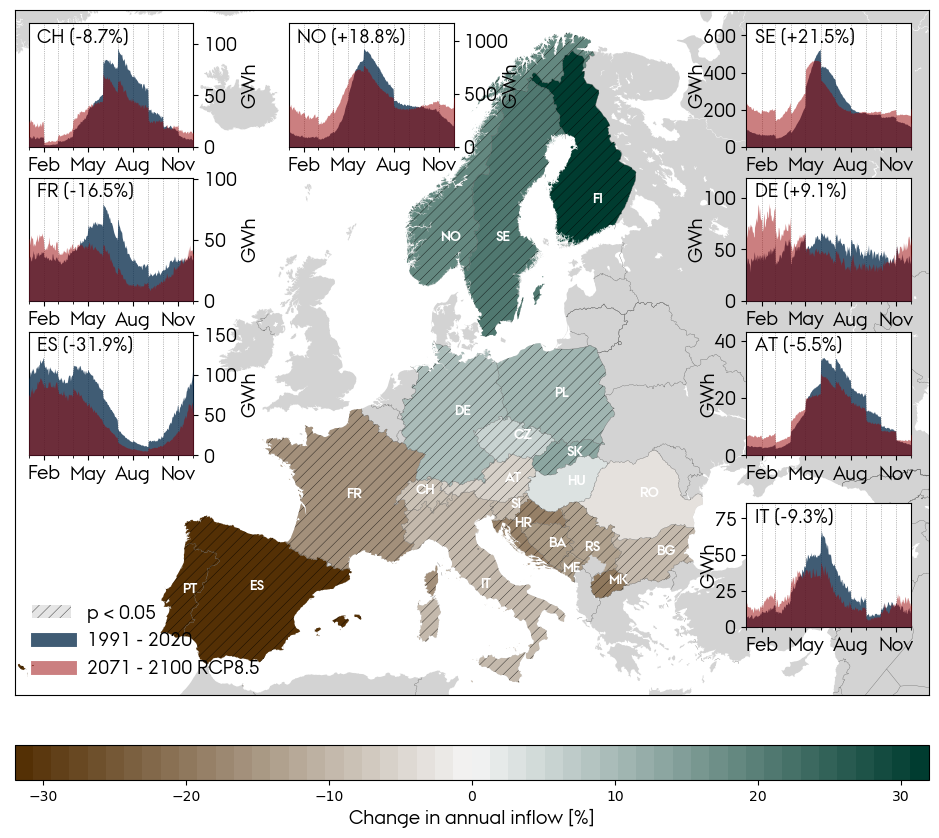}
	\caption{Ensemble mean relative change in annual inflow and change in seasonal inflow profile for RCP8.5. Dashed patterns indicate results that are statistically significant (p<0.05). For RCP4.5, see Figure \ref{fig:inflow_ensemble_mean_rcp45}.}
	\label{fig:inflow_ensemble_mean}
\end{figure}

The average of the signal from the 10 climate models, i.e. the ensemble mean, is presented in Figure \ref{fig:inflow_ensemble_mean} in which the relative change in the annual inflow is indicated on the map of Europe and the change in the seasonal inflow profile is given for the countries with the largest hydropower capacity. The consequence of climate change is latitude-dependent illustrated by the Northern European countries facing an increase in the annual inflow while the Southern European countries, e.g. France, Austria, Switzerland, and Italy, all encounter a scarcer annual resource available for hydropower production. The most extreme impacts are found for Spain and Portugal, 30\% decrease in annual inflow under the RCP8.5 scenario, the Balkan countries (10-20\% decrease) and the Nordic countries (20-30\% increase).
All countries except Spain are projected to encounter increased available hydro resources during the winter since less water is stored as ice at high altitudes, but less during spring and summer. Spain does not show any change in the shape of the seasonal profile but it suffers from a reduced inflow throughout the year.\\

\begin{figure}[h!]
	\centering
	\includegraphics[width=\textwidth]{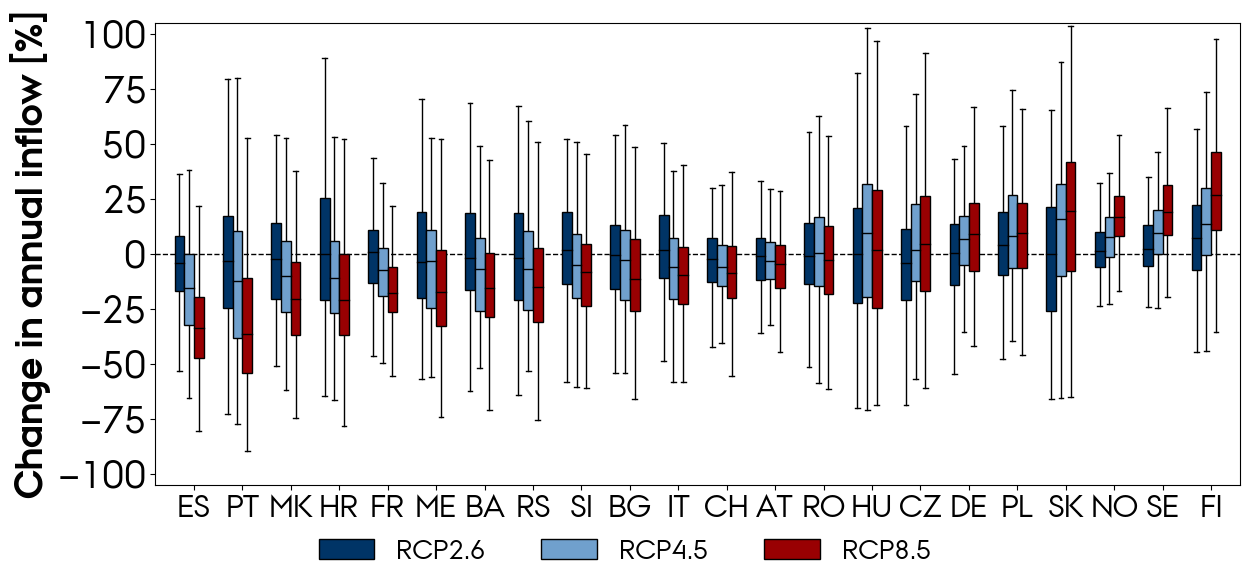}
	\caption{Box and whisker plot of changes in annual inflow at the EOC period (2071-2100) relative to the BOC (1991-2020) for 22 European countries at the three scenarios, RCP2.6,RCP4.5, and RCP8.5.}
	\label{fig:rcp}
\end{figure}

Thankfully, the RCP8.5 scenario has a low probability, given the current emissions trends \parencite{Hausfather2020}, but it is useful to look at it since it provides a more obvious signal. The RCP4.5 scenario is, on the other hand, considered a likely scenario with current mitigation policies \parencite{Hausfather2020}. Figure \ref{fig:rcp} presents the box and whisker plots of the relative change in annual inflow for the 22 countries at the three emissions scenarios (each plot represents the distribution of 6x30, 8x30, and 10x30 model years for RCP2.6, RCP4.5, and RCP8.5). When looking at the median (and other percentiles shown by box and whiskers), the results are generally consistent in the direction of change across the scenarios. For the RCP4.5 scenario, Spain and Portugal still show a robust 11-14\% decrease in annual inflow at the EOC period, the Balkan countries (except Romania) show a 3-10\% decrease, while the Nordic countries show a 8-14\% increase. For Romania, the impact diverges between a 1\% increase and 3\% reduction across the RCP4.5 and RCP8.5.\\ 

\begin{figure}[h!]
	\centering
	\includegraphics[width=0.95\textwidth]{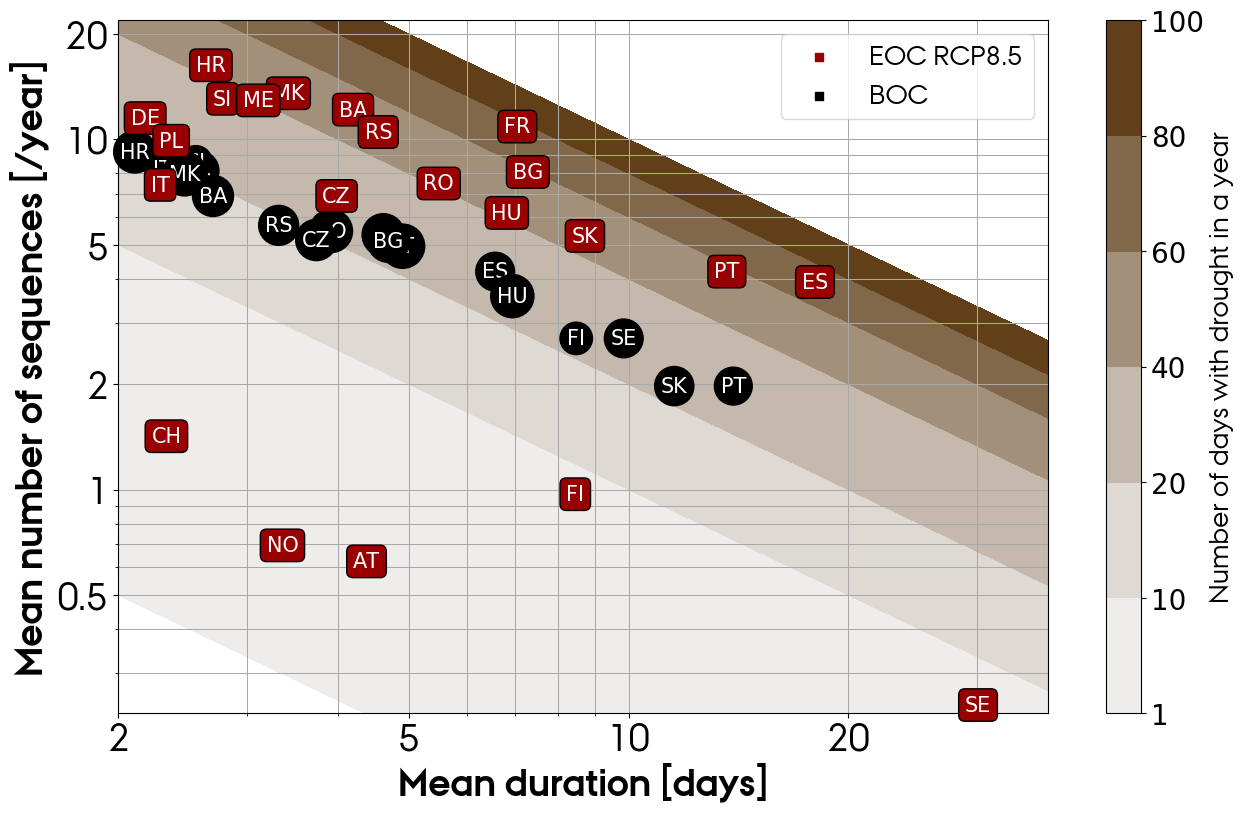}
	\caption{Duration and frequency of droughts, determined as consecutive days with inflow less than the $10^\text{th}$ percentile of inflow at the BOC, evaluated for the BOC and EOC periods under RCP8.5 scenario. For RCP2.6 and RCP4.5, see Figure \ref{fig:extreme_droughts}, and for periods with overflow, see Figure \ref{fig:extreme_overflow}.}
	\label{fig:drought}
\end{figure}

Furthermore, we look at the frequency and severity of extreme events in inflow and how they are impacted by climate change. We define droughts as consecutive days with inflow <10$^\text{th}$ and periods with overflow as >90$^\text{th}$ percentile. Figure \ref{fig:drought} shows the frequency and duration of inflow droughts in the different countries under the RCP8.5 scenario (see Figure \ref{fig:extreme_droughts} for droughts under the RCP2.6 and RCP4.5 scenarios). Similarly, periods with overflow is shown in Figure \ref{fig:extreme_overflow}. As hydropower is expected to balance wind and solar fluctuations, severe droughts lasting several weeks could stress the system operation. Unfortunately, for most countries, climate change is expected to increase the length of drought periods and their frequency. The observed tendency is stronger for Mediterranean countries. In addition, less and shorter periods of overflow is expected in all countries, except Poland, Germany, Czech Republic, Finland, and Slovakia.\\ 


As a final step, we investigate the impacts that the changes in inflow would have on the coupled electricity and heating European energy system. To that end, we compare the cost-optimal net-zero emissions system in 2050, when including the climate projected hydro inflow time series at RCP8.5 (ensemble mean) at the BOC and EOC periods. For those countries where significant changes in hydro resources at the EOC period are expected, we find that the optimal wind and solar capacities vary to compensate them, see Figure \ref{fig:power_capacity_change}. However, the temporal patterns of hydro dispatch, remain almost identical. The mean daily and seasonal patterns are imposed by the need to balance wind and solar while the latter is enabled by the large reservoir capacity in most countries. 

\newpage
\subsection*{CONCLUSION}


In this work, we have investigated the impacts of climate change and high penetration of wind and solar in the operation of reservoir hydropower plants in Europe. First, we showed that balancing renewable generation requires increasing the seasonality of hydropower operation. With the expected large solar capacities in Southern Europe, hydropower plants located in this region are required to mostly operate during winter and nighttime. The perfect foresight, typically assumed in energy models, implies an optimal use of the hydropower balancing potential throughout the year. For most countries, the new dispatch patterns seem possible, based on the historical operation. Higher frequency of large ramps is expected which can lead to reduced lifetime of the hydropower plants. However, for countries with low reservoir energy capacity and summer inflow peaks, e.g. Austria, Switzerland, and Italy, the hydropower dispatch shift from summer to winter, included in the model, may not be feasible. \\

Second, we showed that despite large interannual and significant intermodel variations, we can still detect a change in the inflow caused by climate change. For the high emissions scenario (RCP8.5), using a paired t-test, we found a statistically significant change in annual inflow at the EOC period, relative to BOC, for 20 out of 22 European countries under analysis. The ensemble mean relative change in annual inflow is projected to decrease in Mediterranean and Balkan countries. Spain and Portugal face the largest reduction: 31-32\% for the high emissions scenario and 11-14\% for the mid (RCP4.5). Balkan countries (except Romania) face a reduction in the range of 9-20\% (3-10\%) for the high (mid) emissions scenario. On top of that, both regions will suffer from more frequent and prolonged droughts. Conversely, the ensemble mean relative change in annual inflow is projected to increase in Northern Europe. Annual inflow in Nordic countries increases in the range of 19-32\% (8-14\%) for the high (mid) emissions scenario. The frequency of overflow events is expected to increase for Finland but not for Sweden and Norway. The climate projections show that climate change affect the interannual variability as well. Nordic countries, Germany, Poland, and Slovakia robustly show a larger interannual variability under both mid and high emissions, whereas for Switzerland, Italy, Spain, Bosnia and Herzegovina, and Croatia it is reduced. When designing the future European energy system, climate change impact on hydro resources will require additional wind and solar capacities in southern countries.




\subsection*{SUPPLEMENTAL INFORMATION}
Supplemental Information is found in  \hyperref[sec:supplementary]{Supplemental Materials}.

\subsection*{ACKNOWLEDGEMENT}
This work was authored by E.K.G and M.V., both funded by 
the RE-INVEST project which is supported by the Innovation Fund Denmark under grant number 6154-00022B.

\subsection*{AUTHOR CONTRIBUTIONS}
Conceptualisation, M.V. and E.K.G.; project administration, visualization, and \newline
writing$-$original draft, E.K.G.; writing$-$review \& editing, E.K.G. and M.V.

\subsection*{DECLARATION OF INTERESTS}
The authors declare no competing interests.

\newgeometry{left=20mm,right=20mm,top=20mm,total={170mm,257mm}}

\subsection*{REFERENCES}
\begin{multicols}{3}
	\printbibliography[heading=none]
\end{multicols}

\label{p:lastpage}

\restoregeometry

%% file: Chapters/supplementary.tex
\beginsupplement
\chapter*{Supplementary material}\label{sec:supplementary}

\begin{refsection}

\vspace{-1 cm}
\section*{S1 Methods}\label{s:methods}

\subsection*{Upstream determination}

Surface runoff is the excess water from precipitation, i.e. rain and snow, or meltwater from snow or glaciers that does not enter the soil (infiltration) or evaporates into the atmosphere. The runoff movement is actuated by gravity, and the flow direction is determined by the topography, i.e. the shape of the land surface. The runoff will eventually reach a river, a lake, a pond etc. and be collected in here. A drainage basin (referred to as \textit{basin}) is the area of land that drains all flow of water, e.g. runoff, river streams, groundwater etc., to a common outlet, e.g. a river, reservoir, the ocean etc. \cite{USGS}. The basins can be delineated from topographic maps, using the Pfafstetter coding scheme from Verdin and Verdin \cite{Verdin1999}. The Pfafstetter code itself indicates the number of subdivisions and the basin type. If the last digit is odd the basin is an \textit{interbassin} meaning  that the main stem of the river runs through it. If the last digit is even, the basin is a \textit{tributary basin} and only has tributaries connecting to the river. Finally, if the last digit is zero, it is an isolated \textit{internal basin} that does not have any river connections with surrounding basins. Increasing the level, increases the resolution, thus the size of each basin is reduced. By this approach, one digit is added to the Pfafstetter code, e.g. a level 1 code has one digit, level 2 has two digits, etc. The amount of runoff that is caught by the river does not only depend on the runoff in one single subbasin, but does, due to the interconnection, also depend on the runoff collected by the river-connecting basins. The basin area that contributes to the runoff in a reservoir, is termed as the \textit{upstream area}. 

\begin{figure}[ht]
	\centering
	\hspace*{-0.5 cm}
	\includegraphics[width=10cm]{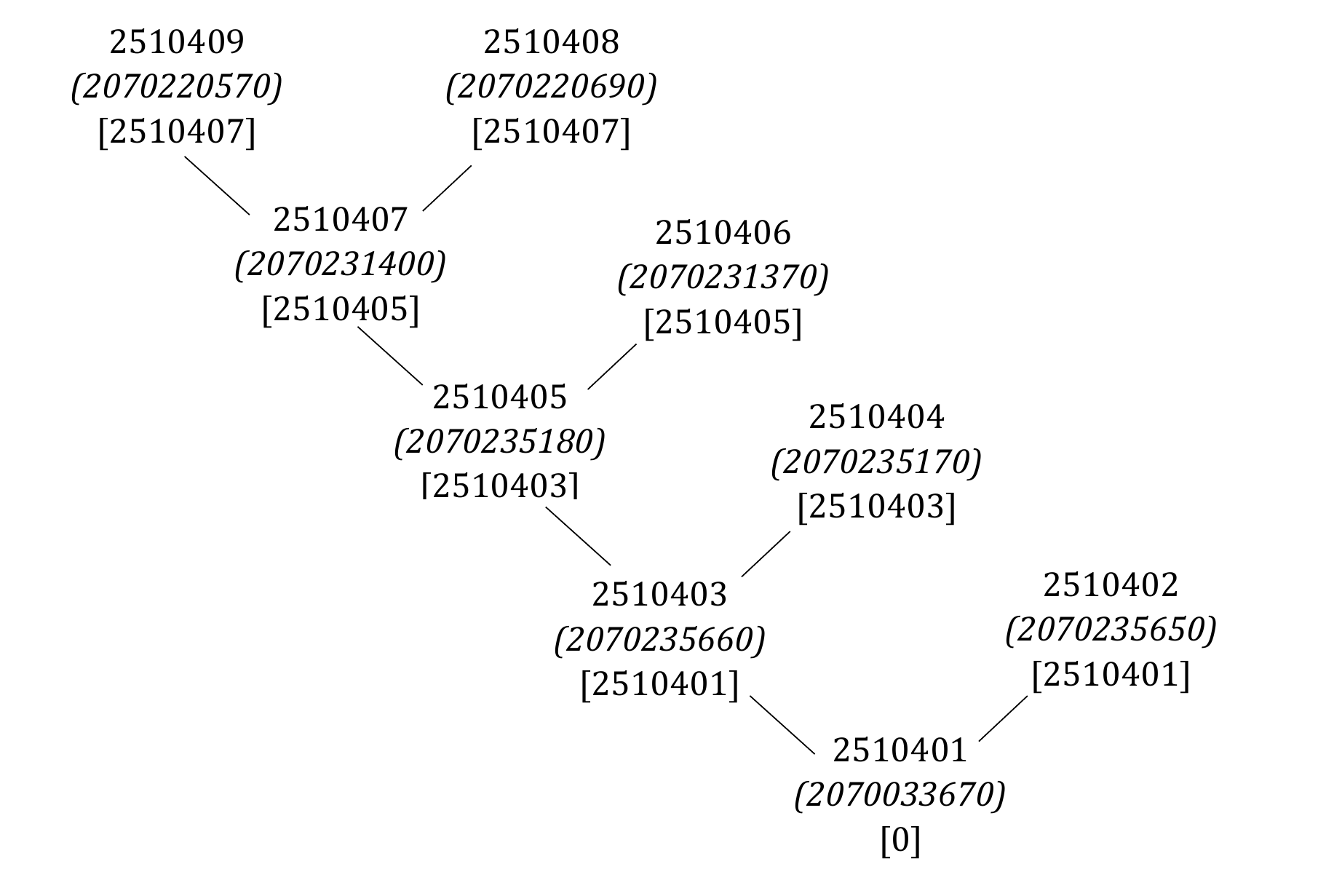}
	\caption{Tree of the level 7 basins connection in one river system in Norway. Numbers are Pfafstetter codes representing each subbasin. In paranthesis the unique basin identifier from the hydroBASINS database is given, and square brackets indicate the Pfafstetter code of the basin into where the river flows, i.e. the next downstream basin, with zero indicating the mouth of the river.}	
	\label{fig:pfafstetter_network}
\end{figure}

\begin{figure}[ht!]
	\centering
	\hspace*{-0.5 cm}
	\includegraphics[width=13cm]{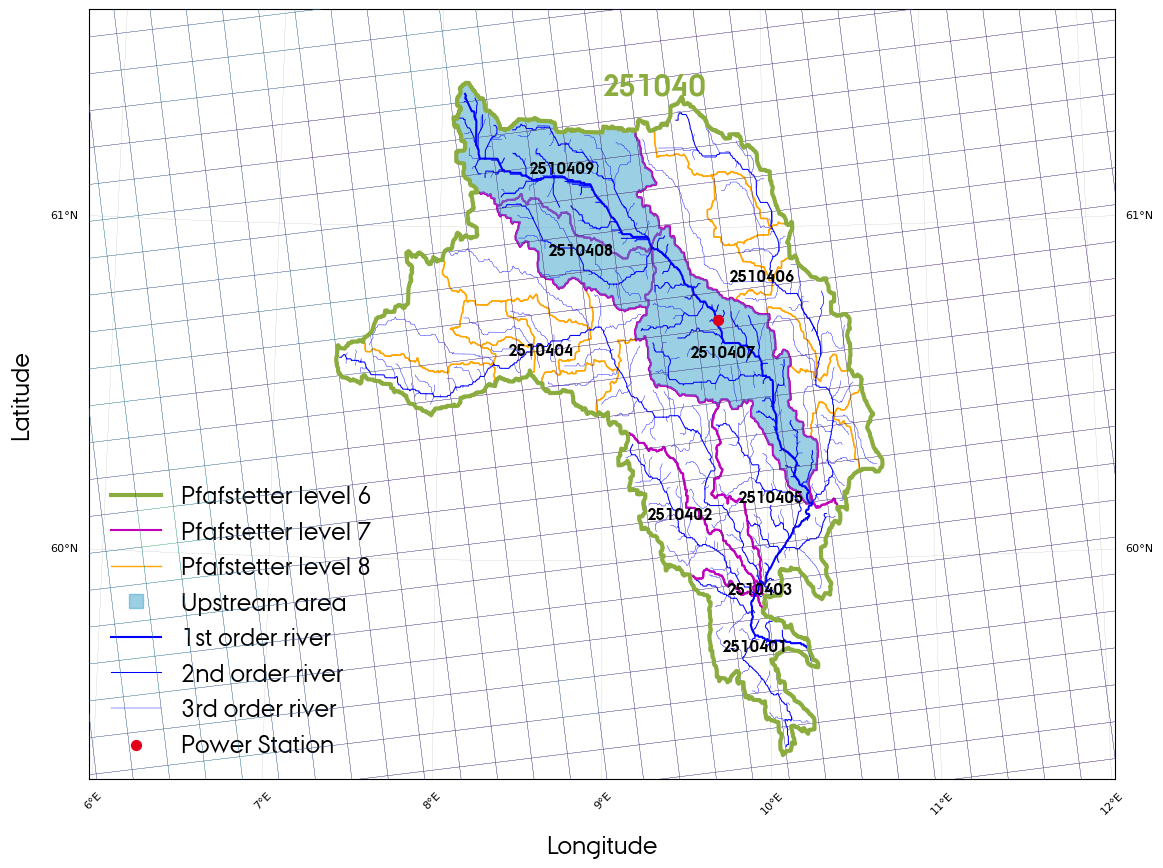}
	\caption{Upstream area depiction of one power plant in Norway, using a level 7 basin delineation, with an overlay of the EURO-CORDEX EUR-11 grid.}	
	\label{fig:upstream_areas_eid}
\end{figure}

Figure \ref{fig:upstream_areas_eid} considers one particular power station located in basin 2510407, in Southern Norway. By looking at the network in Figure \ref{fig:pfafstetter_network}, it is known that basin 2510408 (tributary) and 2510409 (interbasin) contribute to the flow in basin 2510407, and they all constitute the upstream area for the reservoirs in this specific basin. The upstream determination was in this case based on a Pfafstetter level 7 delineation.\\ 

\begin{figure}[ht!]
	\centering
	\hspace*{-0.5 cm}
	\includegraphics[width=12cm]{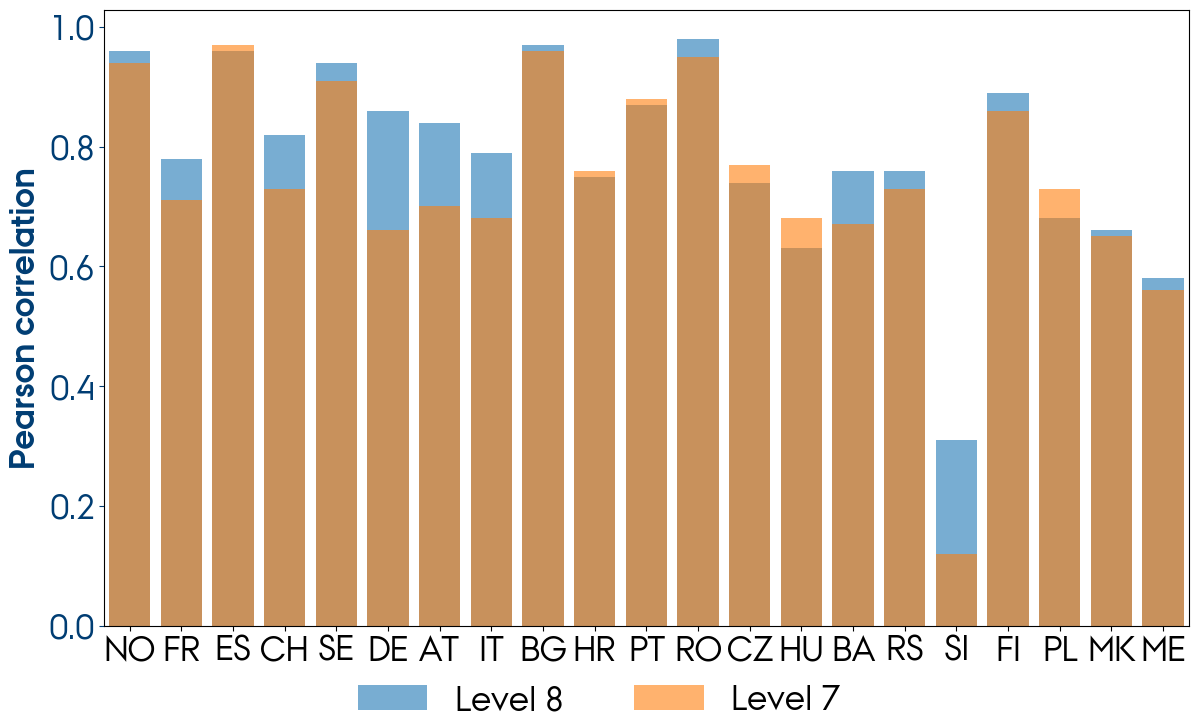}
	\caption{Pearson correlation between modelled (MPI-ESM-LR-RCA4) and historical inflow at Pfafstetter level 7 and 8 basins delineation. Slovakia is excluded from this figure since it is negative in both cases.}	
	\label{fig:basins_level_evaluation}
\end{figure}

As Figure \ref{fig:upstream_areas_eid} also shows, the resolution of the delineation is refined with Pfafstetter level 8. Liu et al. \parencite{Liu2019} applied level 7 at a $0.3^\circ \times 0.3^\circ$ runoff grid. In this work, the runoff data is acuired at a $0.11^\circ \times 0.11^\circ$. In line with this refinement, we determine the upstream basins based on a level 8 basins delineation. A preliminary study shows that in 15 countries countries, level 8 leads to a higher correlation between modelled and historical inflow, shown in Figure \ref{fig:basins_level_evaluation}. 

\newpage
\subsection*{Runoff aggregation}
Using the distance between the runoff grid cell and the reservoir, and assuming a flow speed of 1 m/s, the runoff is first aggregated in the basins connected to a reservoir, followed by a country-level aggregation and conversion into a volumetric flow rate. Country-aggregated volumetric inflow is calculated with Eq. \ref{eq:volumetric_inflow}:
\begin{equation}\label{eq:volumetric_inflow}
	Q_c = \sum_b^M \frac{R_{bc} A_{b}}{\Delta t} \,,
\end{equation}
where $b$ is the upstream basin index, $M$ is the number of upstream basins within country $c$, $R_{bc}$ is the aggregated runoff in basin $b$ and country $c$ $[\SI{}{m}]$, $A_{b}$ is the surface area of the basin $[\SI{}{m^2}]$, $\Delta t$ is the temporal resolution of the runoff data $[\SI{}{s}]$, and $Q_c$ is the country-aggregated volumetric inflow $[\SI{}{m^3/s}]$. 





\subsection*{Energy conversion}

\begin{figure}[ht]
	\centering
	\vspace{-1 cm}
	\includegraphics[width=14cm]{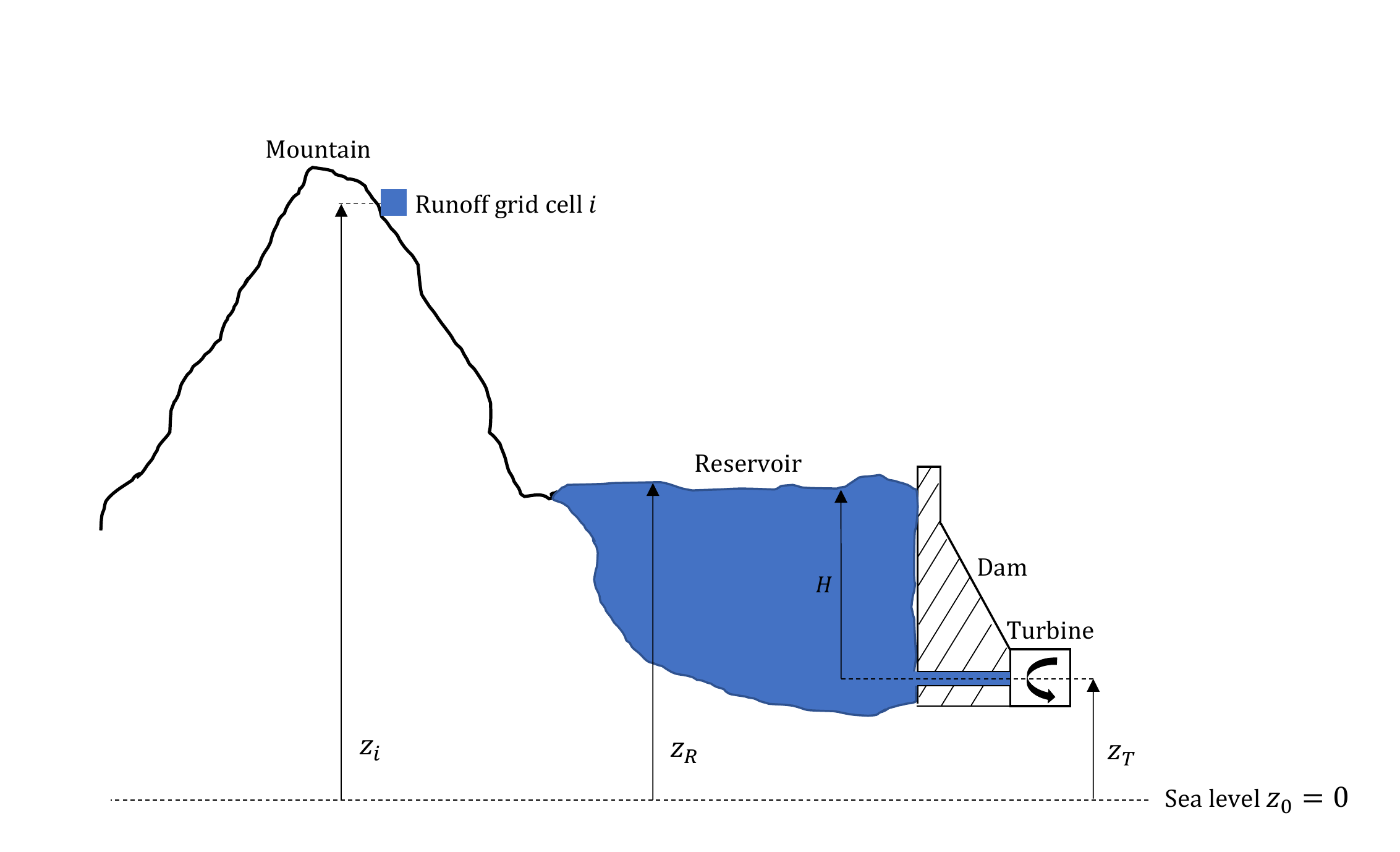}
	\caption{Simplified illustration of one runoff grid cell connected to a hydropower reservoir.}	
	\label{fig:atlite_dam_illustration}
\end{figure}

The aggregated volumetric inflow can be transformed into units of energy to make it comparable with historical inflow. Figure \ref{fig:atlite_dam_illustration} shows a simple illustration of one grid cell $i$ containing runoff at the altitude $z_i$. Assuming that the initial velocity of the runoff is zero, i.e. negligible kinetic energy, the maximum theoretically extractable energy $E^{\text{max}}$ (per time step) of the runoff at altitude $z_i$ by the turbine at altitude $z_T$ is estimated with Eq.  \ref{eq:grid_cell_total_energy}:
\begin{equation}\label{eq:grid_cell_total_energy}
E^{\text{max}} = \eta \rho g Q (z_i - z_T) \Delta t \,,
\end{equation}
where $\eta$ is the turbine efficiency which is assumed to be 1, $g$ is the gravitational acceleration ($\SI{9.81}{m/s^2}$), $\rho$ is the water density ($\SI{1000}{kg/m^3}$), and $Q$ is the flow rate of the runoff. In reality, some of the potential energy of the water in the grid cell is converted into kinetic energy when moving down the terrain and subsequently lost when reaching the reservoir; thus, the available energy $E^{\text{available}}$ is reduced to:
\begin{equation}\label{eq:grid_cell_total_energy_2}
E^{\text{available}} = \rho g Q (z_R - z_T) \Delta t = \rho g Q H \Delta t \,,
\end{equation}
where $z_R$ is the altitude of the hydropower reservoir and $H$ is the (hydraulic) head. Note that for run-of-river power plants the kinetic energy would not be lost, since this type of plant extracts the kinetic energy directly.\\

In Eq. \ref{eq:volumetric_inflow}, we rewrite $R_{bc} = \sum_i^N R_{ibc}$, where $N$ is the number of grid cells within one basin. Furthermore, we approximate the head in Eq. \ref{eq:grid_cell_total_energy_2} by altitude $z_i$, due to the lack of data on reservoir head of every existing hydropower plant, which yields:

\begin{equation}\label{eq:potential_energy_3}
	E_c^{\text{model}} = \rho g \sum^M_b \sum^N_i z_i R_{ibc} A_{b} \,, 
\end{equation}

where $E_c^{\text{model}}$ is the contry-aggregated modelled energy inflow. The head approximation leads to the error $e_z$ caused by the mismatch between the altitude $z_i$ and the reservoir head $H$:
\begin{equation}\label{eq:altitude_mismatch}
	e_z = z_i - H 
\end{equation}


The bias caused by the mismatch will be adjusted in the calibration with the retain factor, described in the following section.

\subsection*{Retain factor}

The retain factor $RF$ quantifies the level of water lost due to evaporation, transpiration, irrigation, or groundwater infiltration, contributing to an energy loss $E^{\text{loss}}$, and is determined for each country based on the seasonal profile obtained from the modelled inflow at the beginning of the century (1991-2020) and the historical observations (2016-2019) by:
\begin{equation}\label{eq:month_country_dep_RF}
RF_{cm} = \frac{\text{Observed inflow}}{\text{Modelled inflow at BOC}} =  \frac{E_{cm}^{\text{hist}}}{E_{cm}^{\text{model}}} \,,
\end{equation}
where index $c$ indicates country (1-22) and $m$ month of an average year (1-12). 

According to the definition presented in Eq. \ref{eq:month_country_dep_RF}, $RF$ corresponds to the fraction of upstream-runoff caught by the hydropower reservoirs, i.e. the \textit{retained} water. The retain factor includes the bias correction for the mismatch between the head and the runoff altitude, $\bar{e_z}$. Evaporation taking place while water is stored in the reservoir can also impact the potential power generation \parencite{Turner2017}. The retain factor also accounts for this.\\

The model performance is, prior to calibration with the retain factor, evaluated on the Pearson correlation betwen modelled and observed inflow:
\begin{equation}\label{eq:Pearson}
r_{\hat{E}E} = \frac{\sum_{m=1}^{12}(\hat{E}_m - \langle \hat{E} \rangle)(E_m - \langle E \rangle)}{\sqrt{\sum_{m=1}^{12}(\hat{E}_m-\langle \hat{E} \rangle ^2})\sqrt{\sum_{m=1}^{12}(E_m - \langle E \rangle)^2}} \,,
\end{equation}
where $\hat{E}_t$ is the modelled inflow at month $m$, $\langle \hat{E} \rangle$ is the modelled mean inflow, $E_m$ is the observed inflow at month $m$, and $\langle E \rangle$ is the observation mean inflow.\\

The modelled inflow for both BOC and EOC is calibrated with the retain factor:
\begin{equation}\label{eq:calibration}
E_{cm}^{\text{calibrated}} = RF_{cm}(E^{\text{loss}},\bar{e_z}) E_{cm}^{\text{model}} = RF_{cm}(E^{\text{loss}},\bar{e_z}) \rho g \sum_b^M \sum_i^N z_i R_{ibc} A_{b}
\end{equation}

By using Eq. \ref{eq:calibration} for the calibration of inflow at EOC, we assume that climate change does not affect the seasonality and magnitude of the retain factor. In reality, the global temperature rise is accompanied by a larger evaporation rate, which increases the energy loss. This is not accounted for in our approach.

\section*{S2 Data}

\subsection*{Historical inflow and dispatch}
\begin{figure}[ht]
	\centering
	\includegraphics[width=\textwidth]{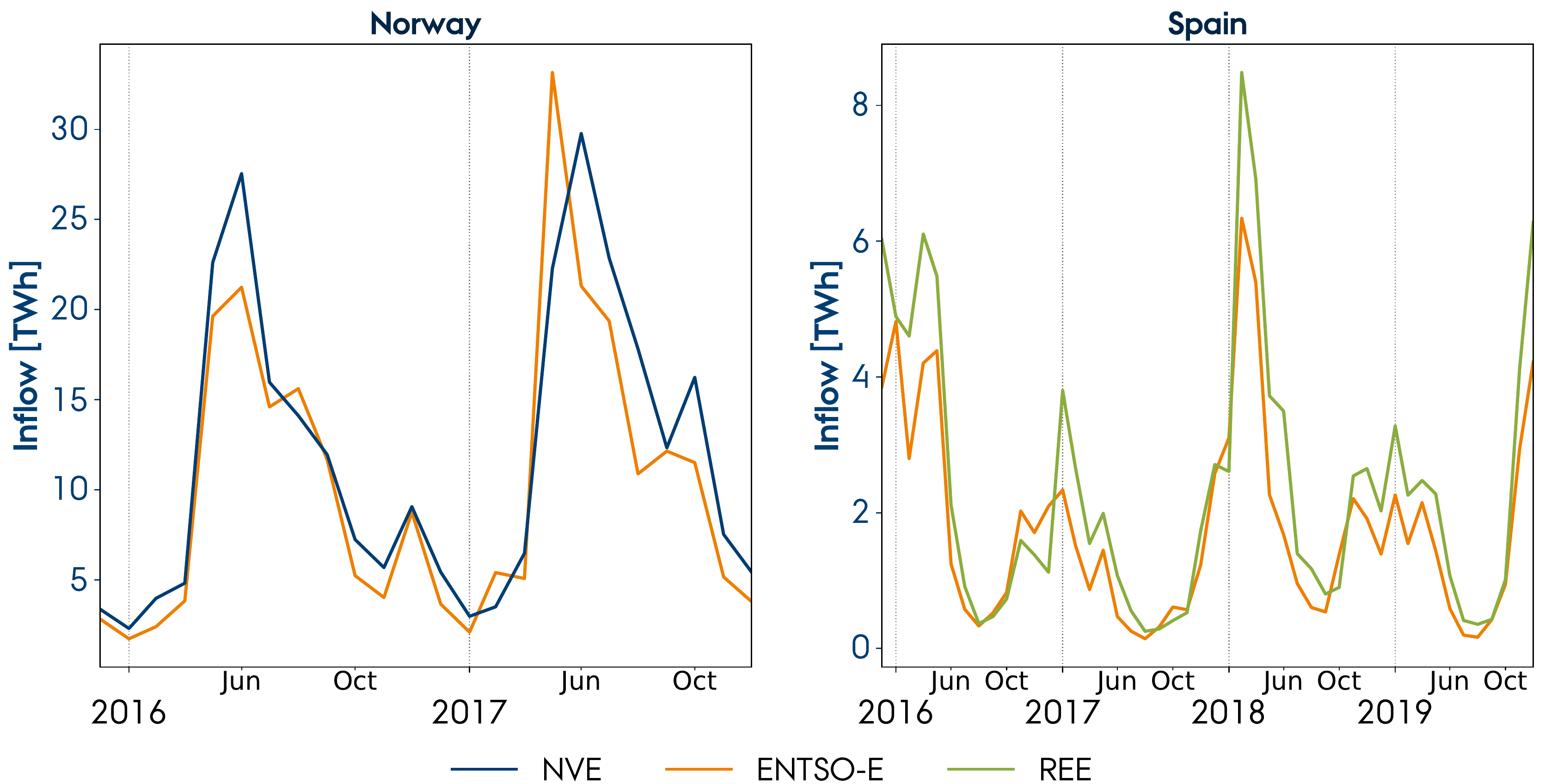}
	\caption{Inflow approximated from electricity generation and reservoir filling level data from ENTSO-E compared with historical inflow from NVE (Norway) and REE (Spain).}
	\label{fig:entso_e_inflow}
\end{figure}

The European Network of Transmission System Operators for Electricity (ENTSO-E) Transparency Platform \parencite{Entso} provides hourly hydropower generation and weekly reservoir filling level. Using an energy balance, the weekly reservoir filling level can be expressed as:
\begin{equation}\label{eq:Reservoir_balance}
	V_{w+1} = V_w + E_w - \sum_{h=1}^{168} G_h - E^{\text{loss}}_w 
\end{equation}
where $V_w$ is the reservoir filling level at week $w$, $V_{w+1}$ the same at week $w+1$, $E_w$ is the energy inflow, $G_h$ is the electricity generation at hour $h$, and $E^{\text{loss}}_w$ is the energy loss. By collecting hourly electricity generation and weekly reservoir filling level at ENTSO-E and neglecting the energy loss, the weekly energy inflow $E_w$ is approximated with Eq. \ref{eq:Reservoir_balance}. The validity of this approximation is evaluated for Norway and Spain by comparing the approximated inflow with historical inflow from The Norwegian Water Resources and Energy Directorate (NVE) \parencite{nve} and Red Eléctrica De España (REE) \parencite{redelectrica}. The comparison is illustrated in Figure \ref{fig:entso_e_inflow} which shows good coherence.\\

\begin{figure}[ht!]
	\centering
	\begin{minipage}{0.95\textwidth}
		\centering
		\subfloat[]
		{
			\includegraphics[width=0.95\textwidth, keepaspectratio]{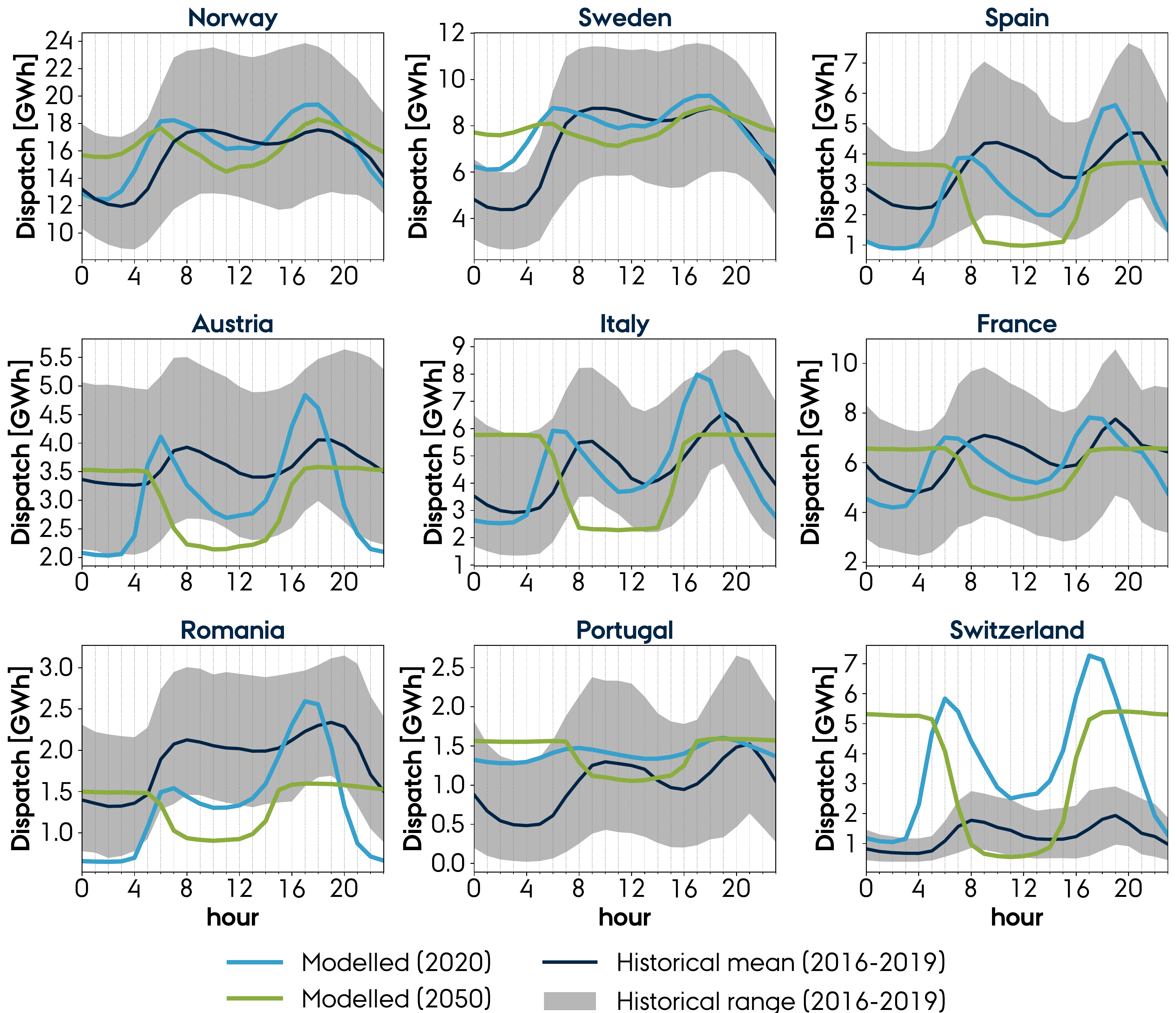}
			\label{fig:intraday_9}
		}
		\\
		\centering
		\subfloat[]
		{
			\includegraphics[width=0.95\textwidth, keepaspectratio]{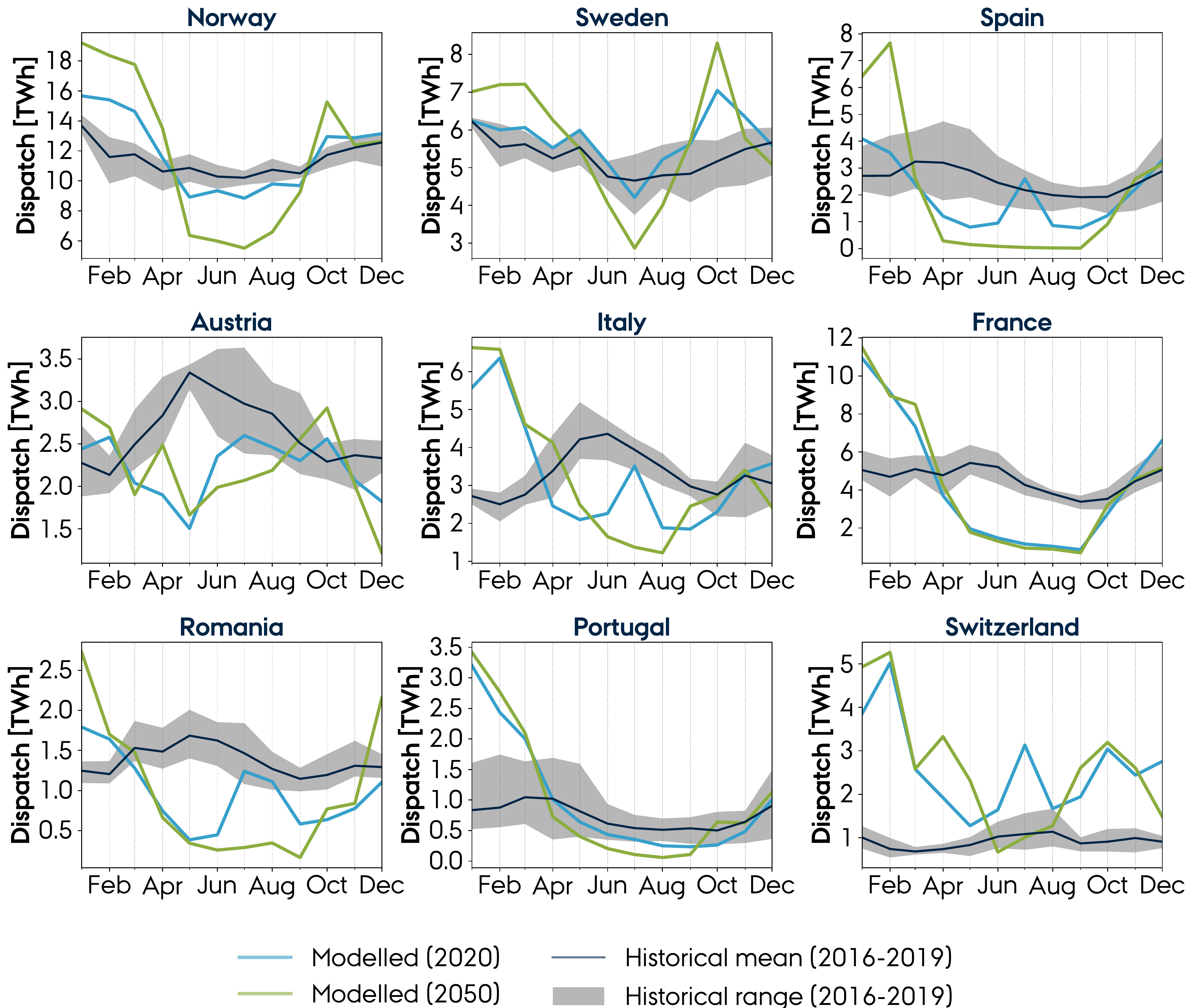}
			\label{fig:seasonal_9}
		}
		\caption{Supplemental to Figure \ref{fig:intraday_dispatch}: Modelled and historical (a) intraday and (b) seasonal operation of hydropower plants.}
		\label{fig:intraday_season_9}
	\end{minipage}
\end{figure}

\begin{figure}[ht!]
	\centering
	\begin{minipage}{\textwidth}
		\centering
		\subfloat[]
		{
			\includegraphics[width=\textwidth, keepaspectratio]{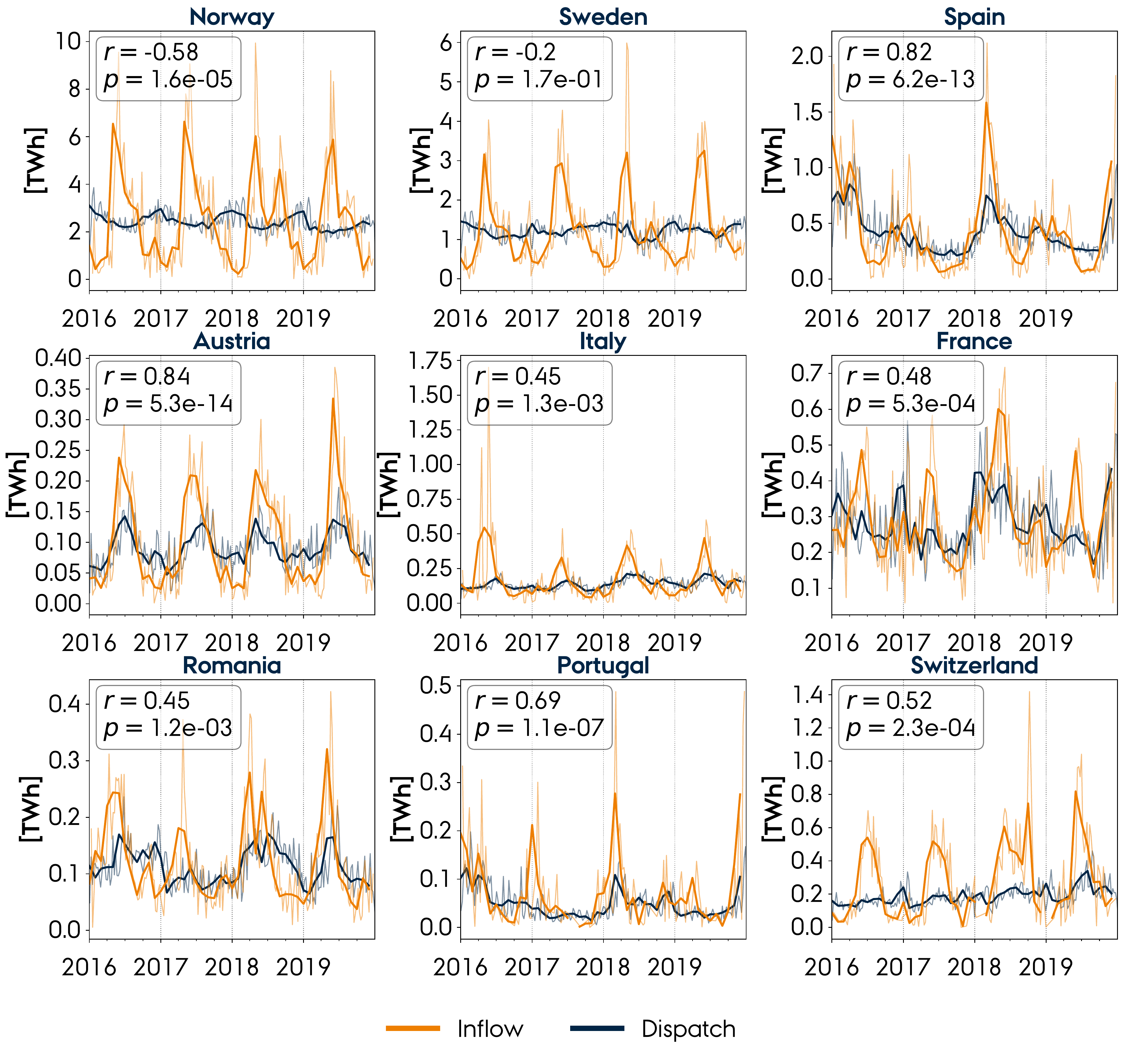}
			\label{fig:d_i_cor_reservoir_ror_9}
		}
		\\
		\centering
		\subfloat[]
		{
			\includegraphics[width=0.7\textwidth, keepaspectratio]{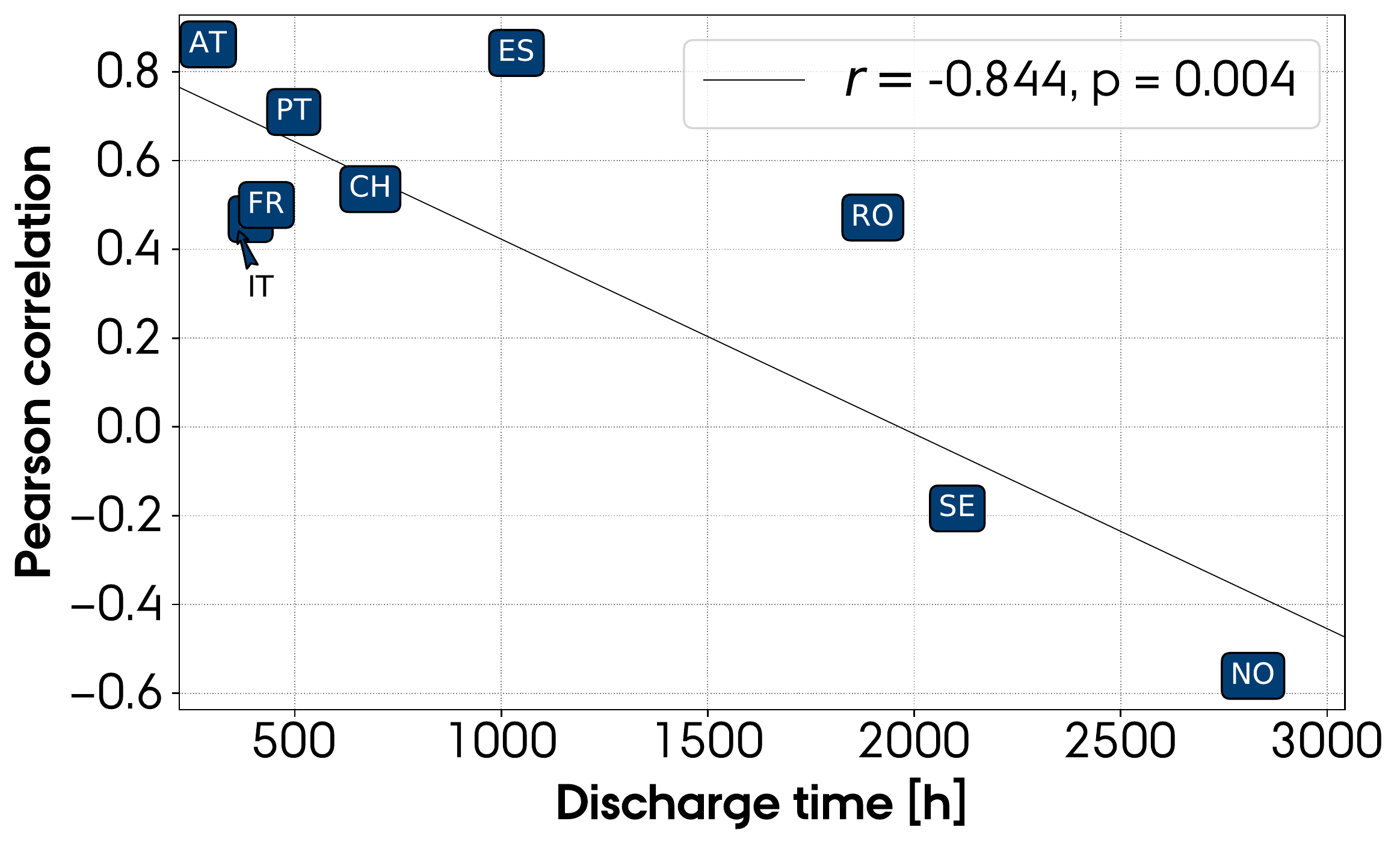}
			\label{fig:dischargetime_pearson_cor}
		}
		\caption{Supplemental to Figure \ref{fig:d_i_cor_reservoir_ror}: (a) Historical hydropower reservoir inflow and dispatch from 2016 to 2019 from ENTSO-E (thin lines indicate weekly values and thick monthly averaged), and (b) inflow-dispatch correlation and discharge time. In (a) Pearson correlation $r$ represents the correlation between inflow and dispatch, and in (b) $r$ represents the correlation between the inflow-dispatch correlation and the discharge time. Furthermore, $p$ is the statistical significance of the Pearson correlation.}
		\label{fig:d_i_cor}
	\end{minipage}
\end{figure}

\begin{figure}[ht]
	\centering
	\includegraphics[width=\textwidth]{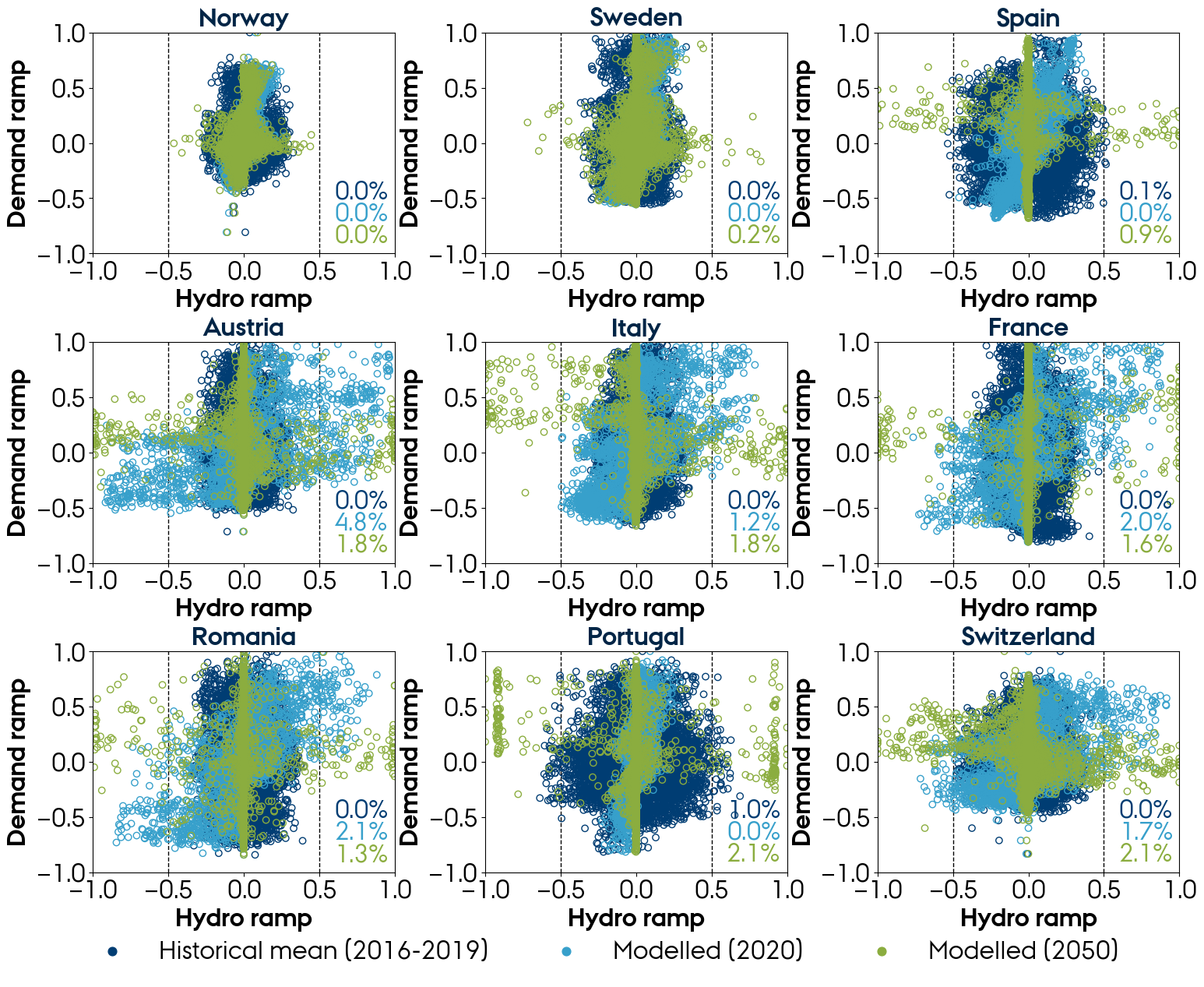}
	\caption{Scatter plot of hourly hydropower ramp rates (x-axis) at given electricity demand ramp rates (y-axis) for the modelled 2020 (cyan), 2050 (green), and historical (dark blue) power production. The percentage of hours in a year at which the absolute value of the ramp rates are above 0.5 is indicated for the three data sets.}
	\label{fig:ramp_rates_9}
\end{figure}

Data from ENTSO-E is available for Austria, Bulgaria, France, Italy, Portugal, Romania, Switzerland, Montenegro, Norway, Spain, and Sweden with a period ranging from 2016 to 2019. Countries such as Norway, Spain, and Sweden have a wider time range of historical inflow available through NVE \parencite{nve} (1958-2017), REE \parencite{redelectrica} (1990-2019), and Swedenergy \parencite{swedenergy} (1980-2019), which are used instead of ENTSO-E. Historical data for the remaining countries which are not available at ENTSO-E, i.e. Bosnia-Herzegovina, Croatia, Czech Republic, Finland, Germany, Poland, Serbia, Slovakia, Slovenia, Hungary, and North Macedonia is provided by Wattsight \parencite{wattsight}, but will not be openly available. For the dispatch intraday and seasonal profile comparison in Figure \ref{fig:intraday_dispatch} and \ref{fig:intraday_season_9}, the historical data is based on hourly reservoir and run-of-river hydropower generation from ENTSO-E. To analyse the dispatch-inflow correlation in Figure \ref{fig:d_i_cor_reservoir_ror} and \ref{fig:d_i_cor_reservoir_ror_9}, and the ramp rate in Figure \ref{fig:ramp_rates_9}, weekly reservoir inflow derived from ENTSO-E reservoir dispatch and filling level is compared with ENTSO-E reservoir dispatch.

\subsection*{Modelled dispatch}
Modelled dispatch is acquired from the hourly-resolved sector-coupled
network model of the European energy system consisting of 30 nodes from Victoria et al. \parencite{Victoria2020} before (2020) and after decarbonisation (2050). A key characteristic of the model is that it assumes a
perfect foresight, which implies that electricity consumption and the inflow is known one year forward. The perfect foresight enables the system to optimize the usage of hydropower, and the water in the reservoirs is thus only released when it is cost-optimal for the system. 

\subsection*{Climate model data}
Climate model data is acquired from regional climate models available at ESGF \parencite{Cordex_database} given with a combination of 5 GCMs (MPI-M-MPI-ESM-LR\parencite{Giorgetta2013},  ICHEC-EC-EARTH\parencite{Hazeleger2012}, CNRM-CERFACS-CM5\parencite{Voldoire2013}, MOHC-HadGEM2-ES\parencite{Collins2011}, and 

NCC-NorESM1-M\parencite{Bentsen2013}) and 2 RCMs (RCA4\parencite{Samuelssson2011} and HIRHAM5\parencite{Christensen2007}). Each climate model is of $0.11^\circ \times 0.11^\circ$ horisontal and daily resolution and data is acquired for the BOC (1991-2020) and EOC (2071-2100) periods. Some climate models are subject to different calendar systems. E.g. NorESM1-M does not include leap years and HadGEM2-ES uses a 360 days per year calendar with 30 days in each month. Furthermore, HadGEM2-ES does not include the year 2100, and, thus, only 29 years will describe the EOC period with this GCM. In these irregular situations, the missing days will be filled out by repeating the data of the previous day, month or year.

\newpage
\subsection*{Location of hydropower plants}

\begin{figure}[ht]
	\centering
	\includegraphics[width=\textwidth]{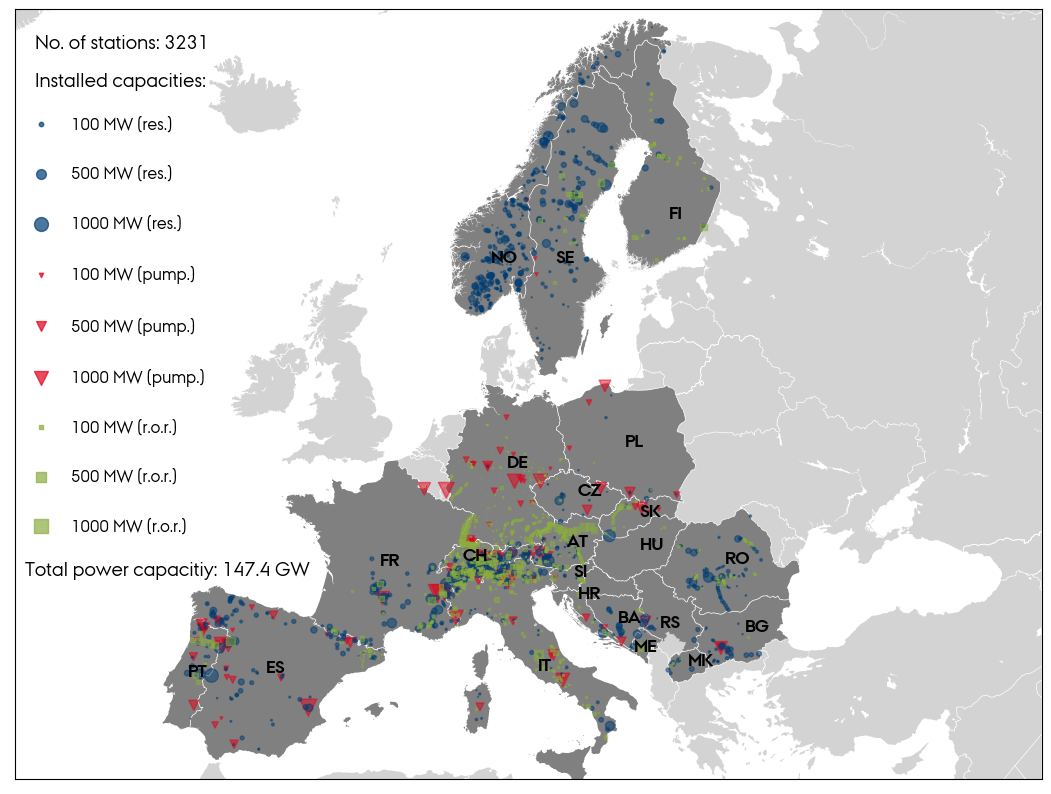}
	\caption{Location and power capacities of reservoirs (res.), pumped-hydro storages (pump.), and run-of-river (r.o.r) power plants from JRC \parencite{jrc}.}
	\label{fig:loc_power_plants}
\end{figure}

Locations of hydropower plants are collected from the JRC Hydropower Database \parencite{jrc}. The location and power capacities from the database for the 22 considered countries are illustrated in Figure \ref{fig:loc_power_plants}. 

\subsection*{Basins delineation}

The boundaries of the watersheds, i.e. the basins delineations, are collected from the HydroBASINS \parencite{Lehner2013} with Pfafstetter level 8.

\newpage
\section*{S3 Results}\label{sec:sup_res}

\newpage



\newpage

\newpage
\begin{figure}[ht]
	\centering
	\includegraphics[width=\textwidth]{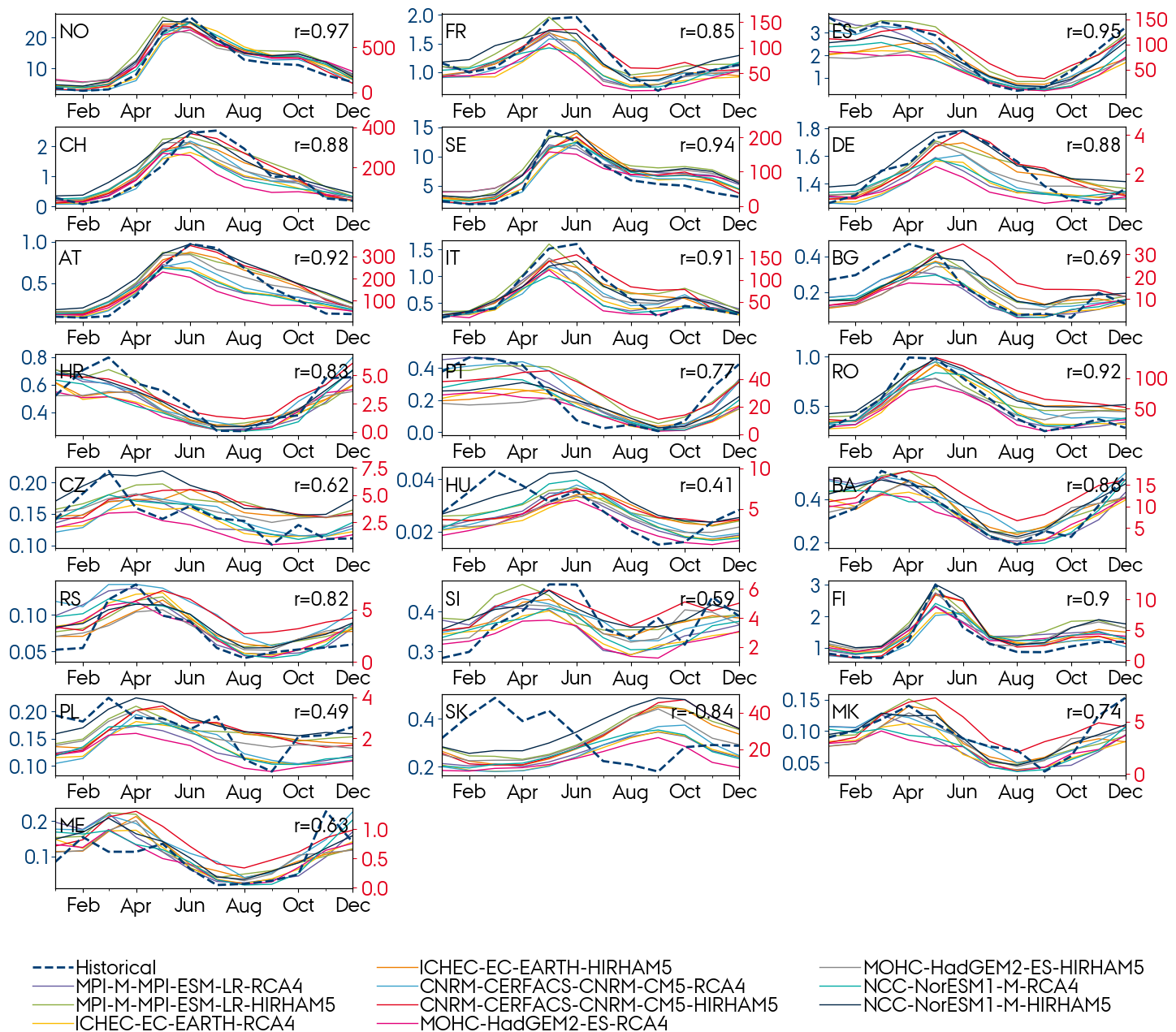}
	\caption{Supplemental to Figure \ref{fig:model_evaluation}:  Comparison of modelled (right axis) and observed (left axis) seasonal inflow in TWh for the 22 European countries. The variable $r$ indicates the mean Pearson correlation obtained for the 10 climate models. The modelled seasonal inflow for Slovakia does not show coherence with the historical observations, illustrated by the climate models consistently predicting peak inflow during fall when it historically has occured during spring, leading to a negative Pearson correlation.}
	\label{fig:model_evaluation_22}
\end{figure}


\newpage
\begin{figure}[ht]
	\centering
	\includegraphics[width=\textwidth]{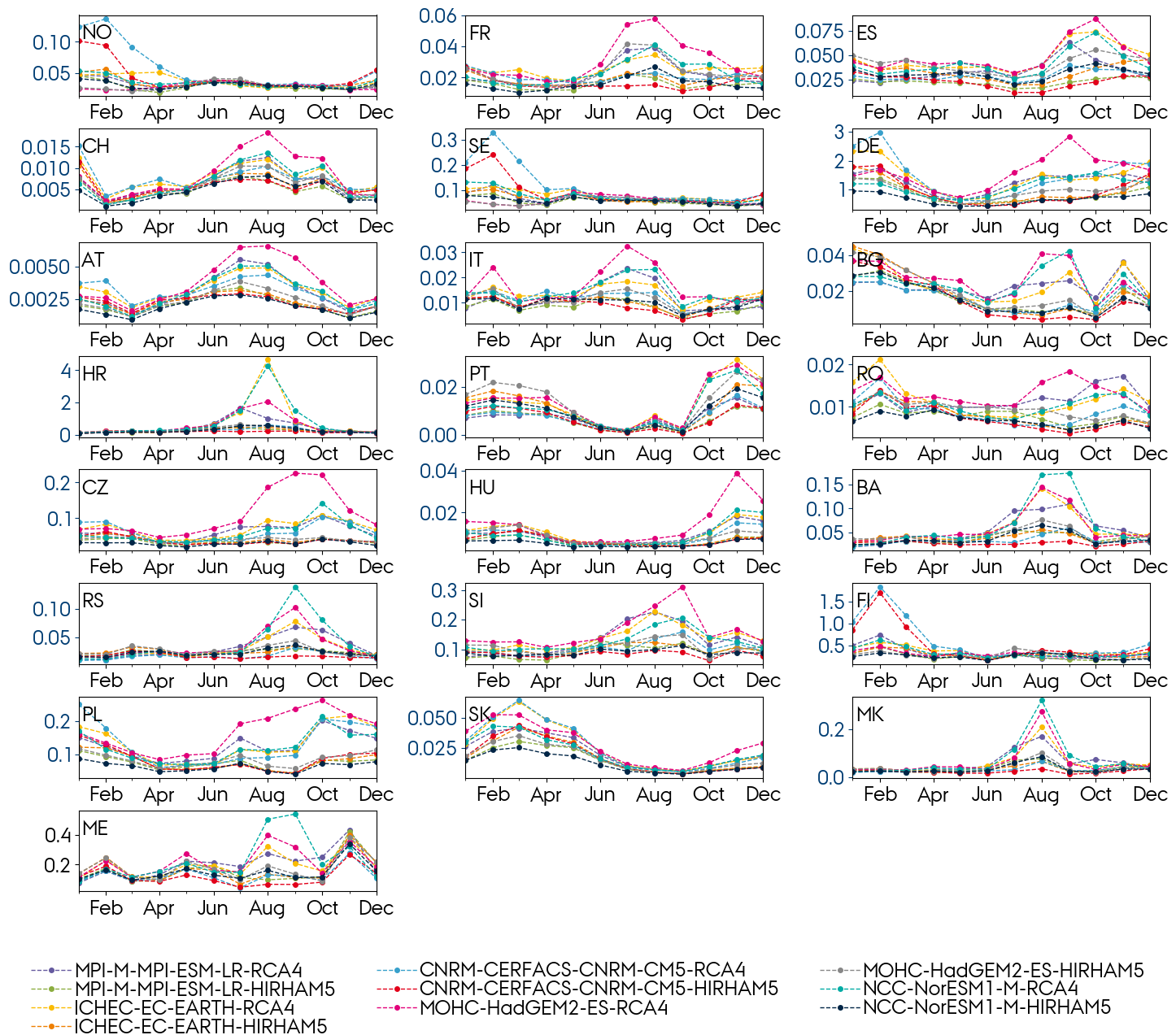}
	\caption{Supplemental to Figure \ref{fig:model_evaluation}: Retain factors for the 22 European countries obtained with the 10 climate models.}
	\label{fig:retain_factors_22}
\end{figure}





\newpage
\begin{figure}[ht]
	\centering
	\includegraphics[width=\textwidth]{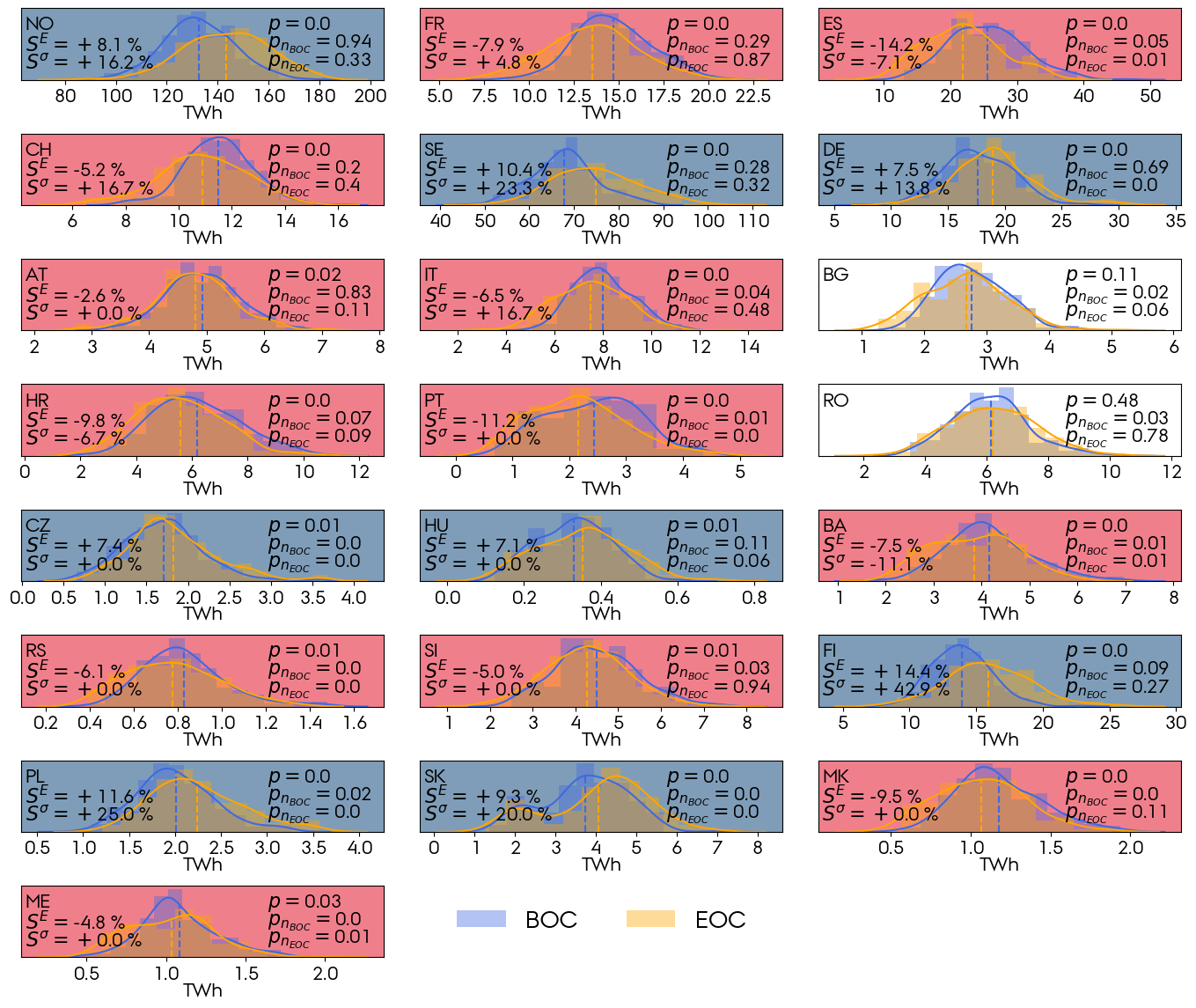}
	\caption{Supplemental to Figure \ref{fig:t_test}: Ensemble distributions of the annual inflow at the BOC and EOC 30-years periods for the RCP4.5 scenario. Dashed lines indicate the mean values of the distributions. The sets are normally distributed if $p_n>0.05$ based on a Shapiro-Wilk test. For the countries with a statistically significant change ($p<0.05$), a blue (red) shade indicates an increase (decrease) in the annual inflow. $S^E$ and $S^\sigma$ correspond to the relative change in annual inflow and interannual variability caused by climate change.}
	\label{fig:t_test_45}
\end{figure}

\newpage
\begin{figure}[ht]
	\centering
	\includegraphics[width=\textwidth]{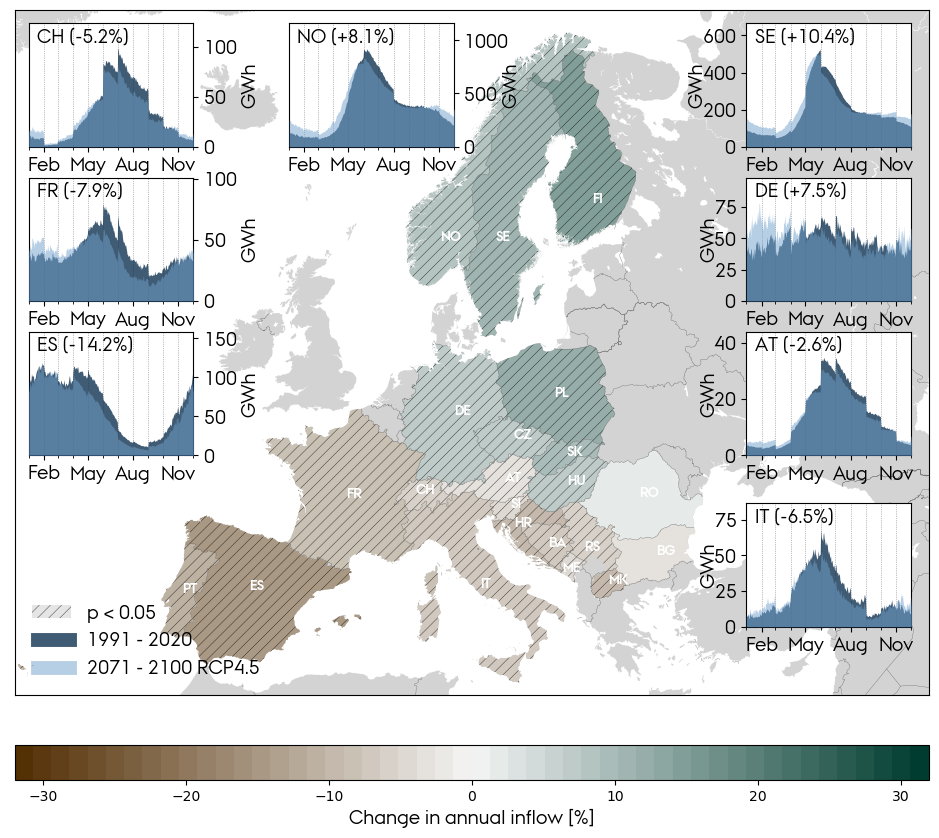}
	\caption{Supplemental to Figure \ref{fig:inflow_ensemble_mean}: Ensemble mean relative change in annual inflow and change in seasonal inflow profile for RCP4.5. Dashed patterns indicate results that are statistically significant (p<0.05)}
	\label{fig:inflow_ensemble_mean_rcp45}
\end{figure}

\newpage
\begin{figure}[ht!]
	\centering
	\begin{minipage}{\textwidth}
		\centering
		\subfloat[]
		{
			\includegraphics[width=0.8\textwidth, keepaspectratio]{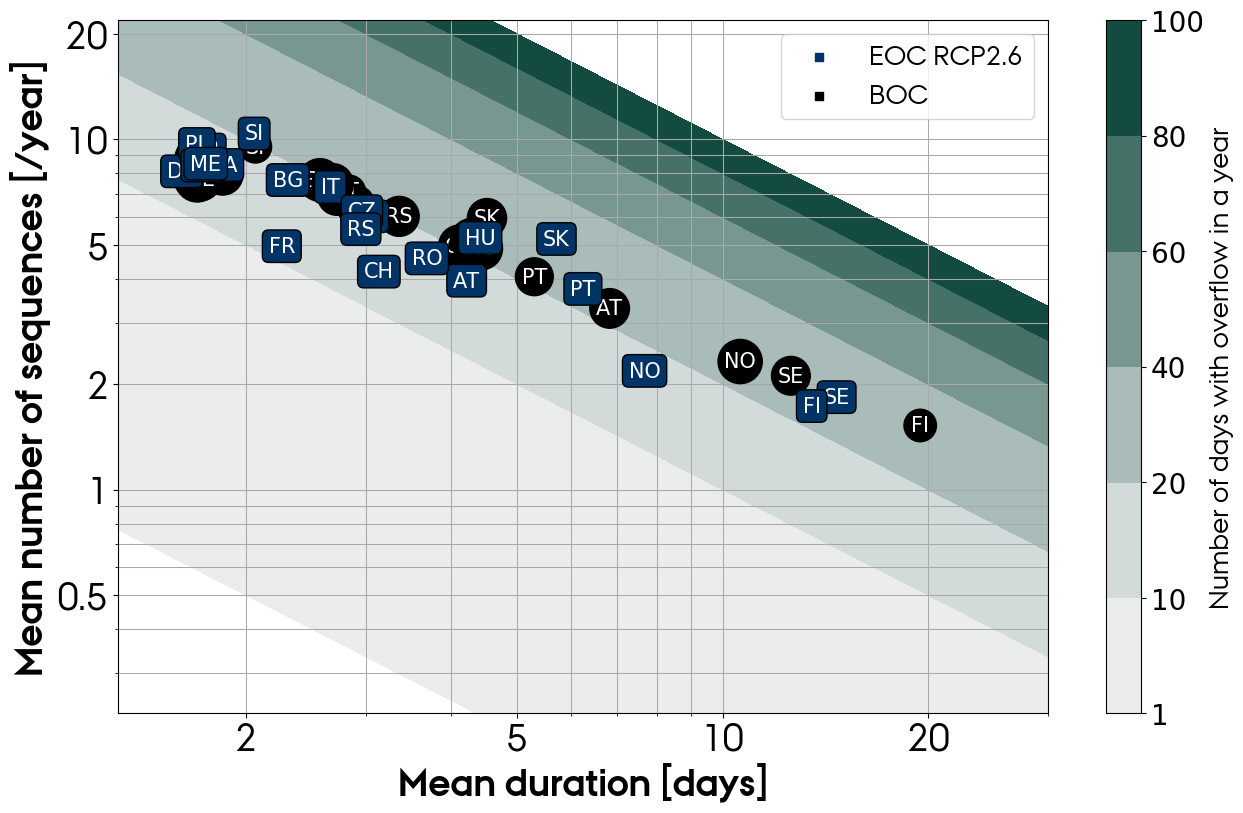}
		}
		\\
		\centering
		\subfloat[]
		{
			\includegraphics[width=0.8\textwidth, keepaspectratio]{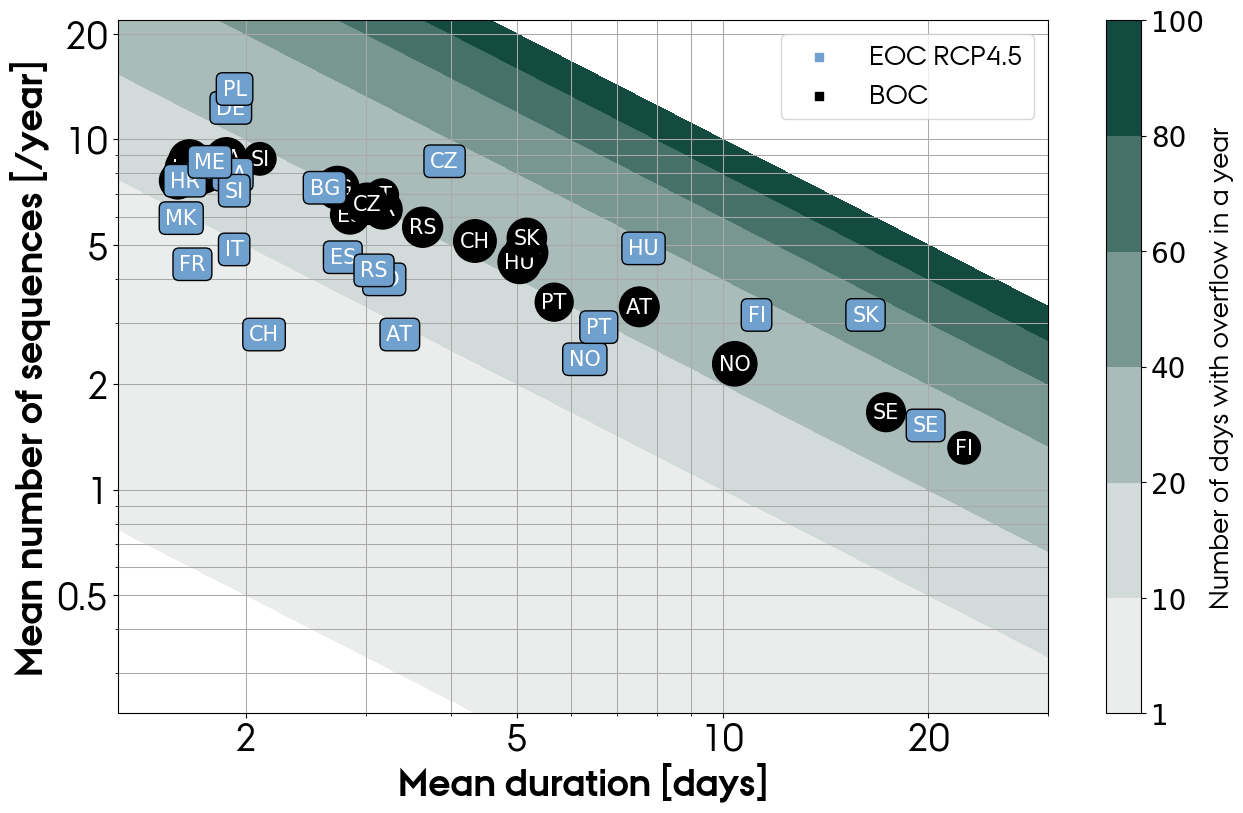}
		}
		\\
		\centering
		\subfloat[]
		{
			\includegraphics[width=0.8\textwidth, keepaspectratio]{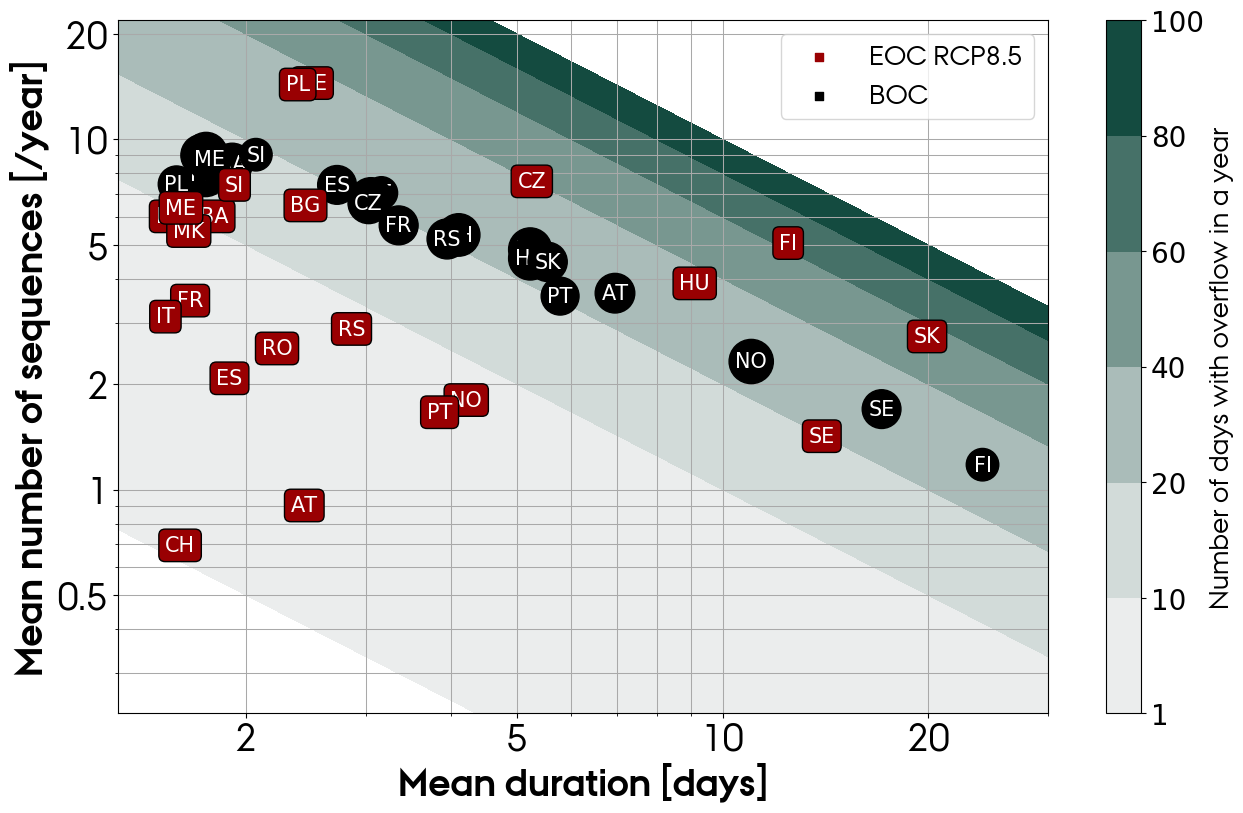}
		}
		\caption{Supplemental to Figure \ref{fig:drought}: Duration and frequency of overflow periods determined as consecutive days with inflow larger than the $90^\text{th}$ percentile of the BOC period, evaluated for the BOC and EOC periods at the (a) RCP2.6, (b) RCP4.5, and (c) RCP8.5 scenario.}
		\label{fig:extreme_overflow}
	\end{minipage}
\end{figure}

\newpage
\begin{figure}[ht!]
	\centering
	\begin{minipage}{\textwidth}
		\centering
		\subfloat[]
		{
			\includegraphics[width=0.8\textwidth, keepaspectratio]{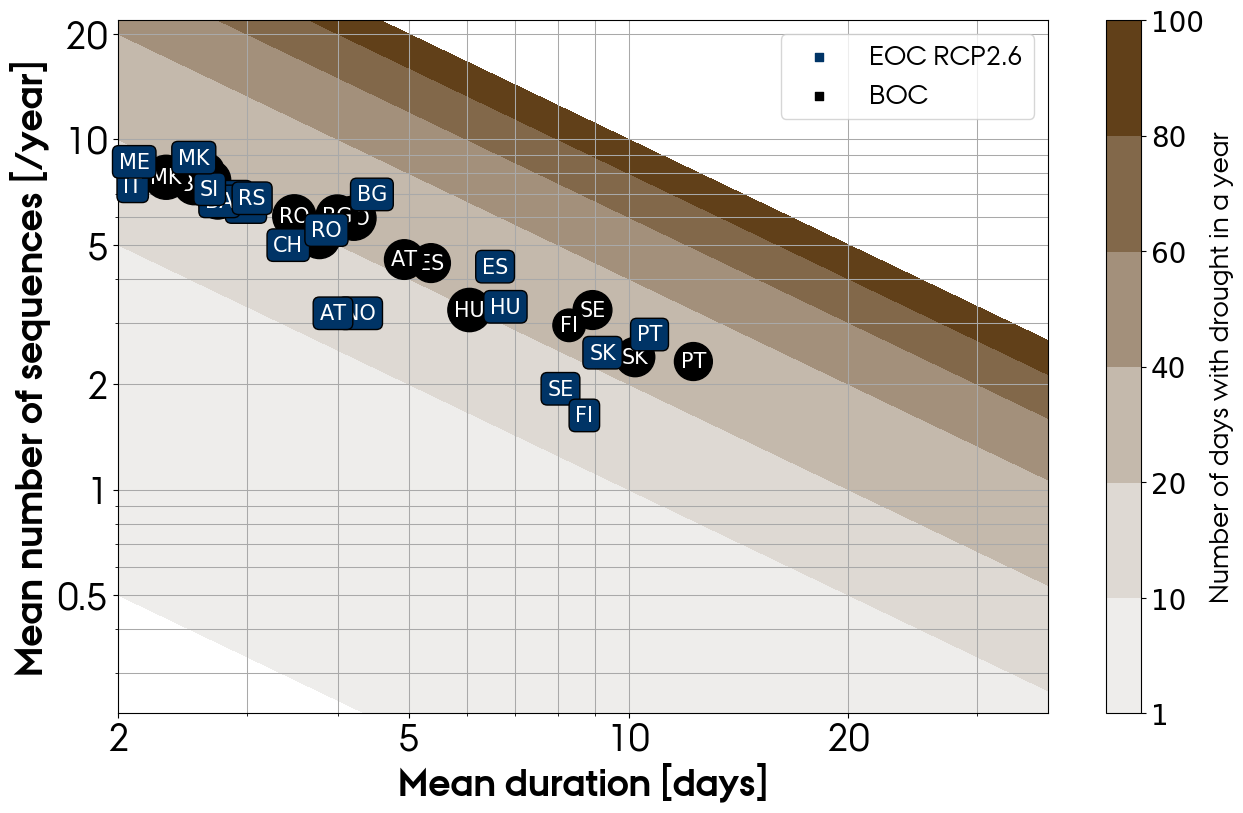}
		}
		\\
		\centering
		\subfloat[]
		{
			\includegraphics[width=0.8\textwidth, keepaspectratio]{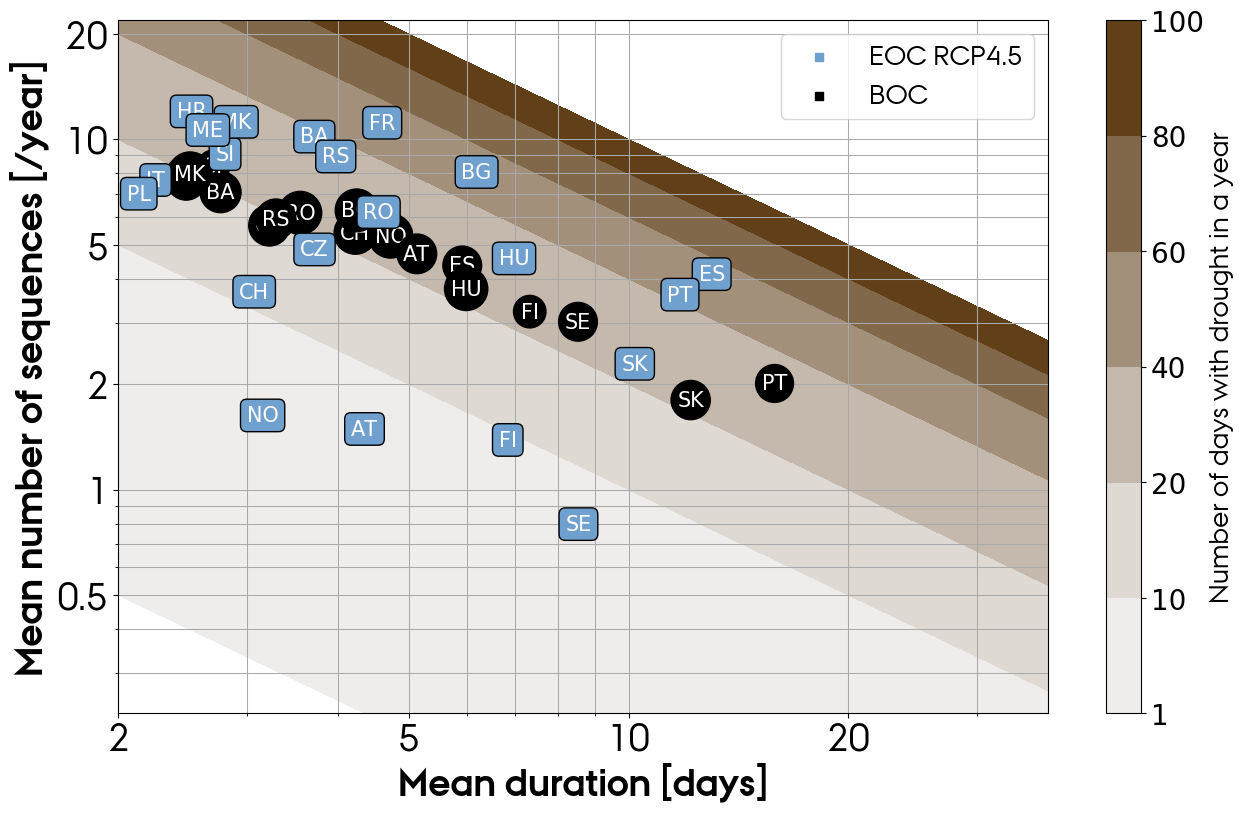}
		}
		\\
		\centering
		\subfloat[]
		{
			\includegraphics[width=0.8\textwidth, keepaspectratio]{Billeder/Extreme_events_droughts_85}
		}
		\caption{Supplemental to Figure \ref{fig:drought}: Duration and frequency of drought periods determined as consecutive days with inflow less than the $10^\text{th}$ percentile of the BOC period, evaluated for the BOC and EOC periods at the (a) RCP2.6, (b) RCP4.5, and (c) RCP8.5 scenario.}
		\label{fig:extreme_droughts}
	\end{minipage}
\end{figure}



\newpage
\begin{figure}[ht!]
	\centering
	\begin{minipage}{\textwidth}
		\centering
		\subfloat[]
		{
			\includegraphics[width=\textwidth, keepaspectratio]{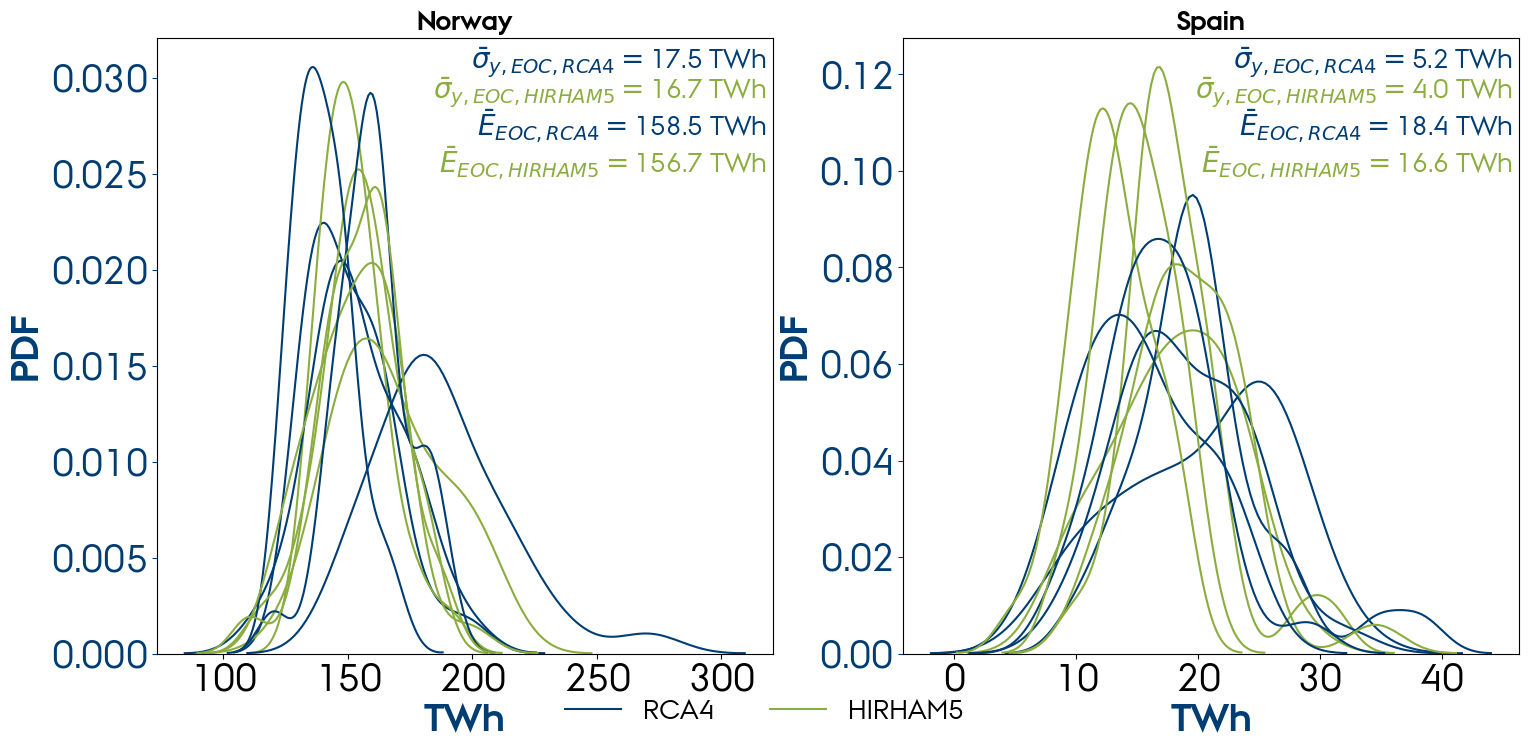}
			\label{fig:a1}
		}
		\\
		\centering
		\subfloat[]
		{
			\includegraphics[width=\textwidth, keepaspectratio]{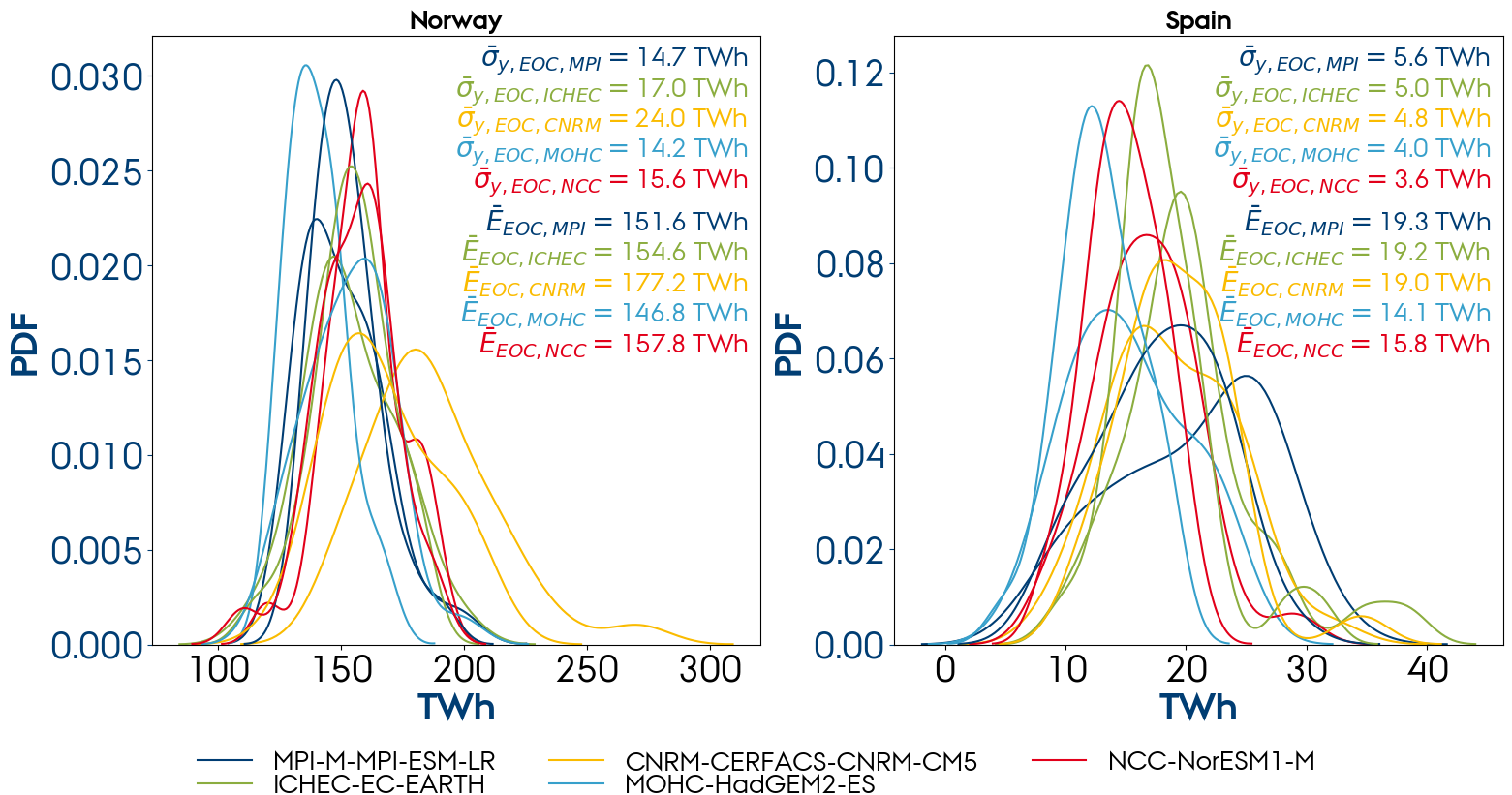}
			\label{fig:c1}
		}
		\caption{(a) Inter-RCM and (b) inter-GCM variability of the annual inflow at the end of the century obtained with the 10 climate models. $\bar{\sigma}_{y}$ is the mean interannual variability (standard deviation) and $\bar{E}$ is the mean annual inflow.}
		\label{fig:intermodel}
	\end{minipage}
\end{figure}

\newpage
\begin{figure}[ht]
	\centering
	\includegraphics[width=\textwidth]{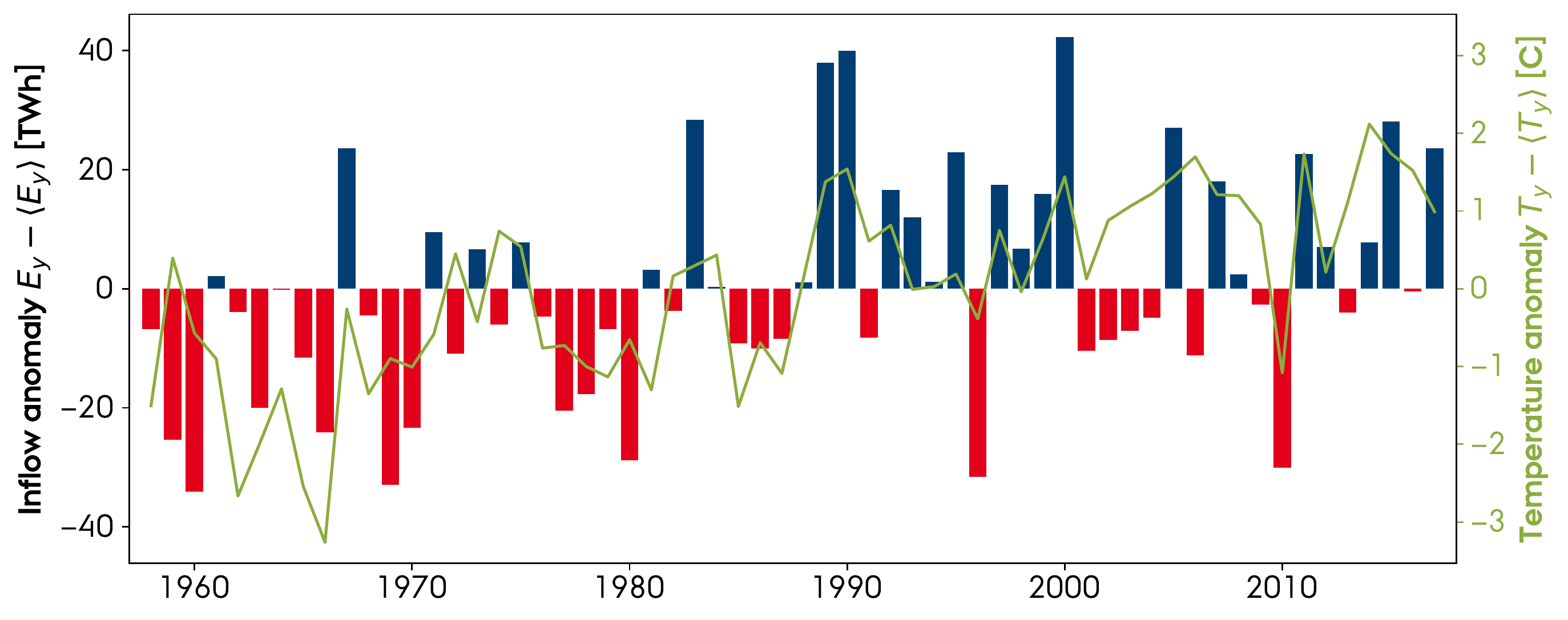}
	\caption{Annual inflow relative to historical mean (left axis) in Norway from 1958 to 2017 \parencite{nve}. The red (blue) bar indicates a year with inflow less (larger) than the historical mean, and the green line plot indicates the air temperature \parencite{era5} relative to the historical mean (right axis).}
	\label{fig:historical_cc_no}
\end{figure}

\newpage
\begin{figure}[ht]
	\centering
	\includegraphics[width=0.95\textwidth]{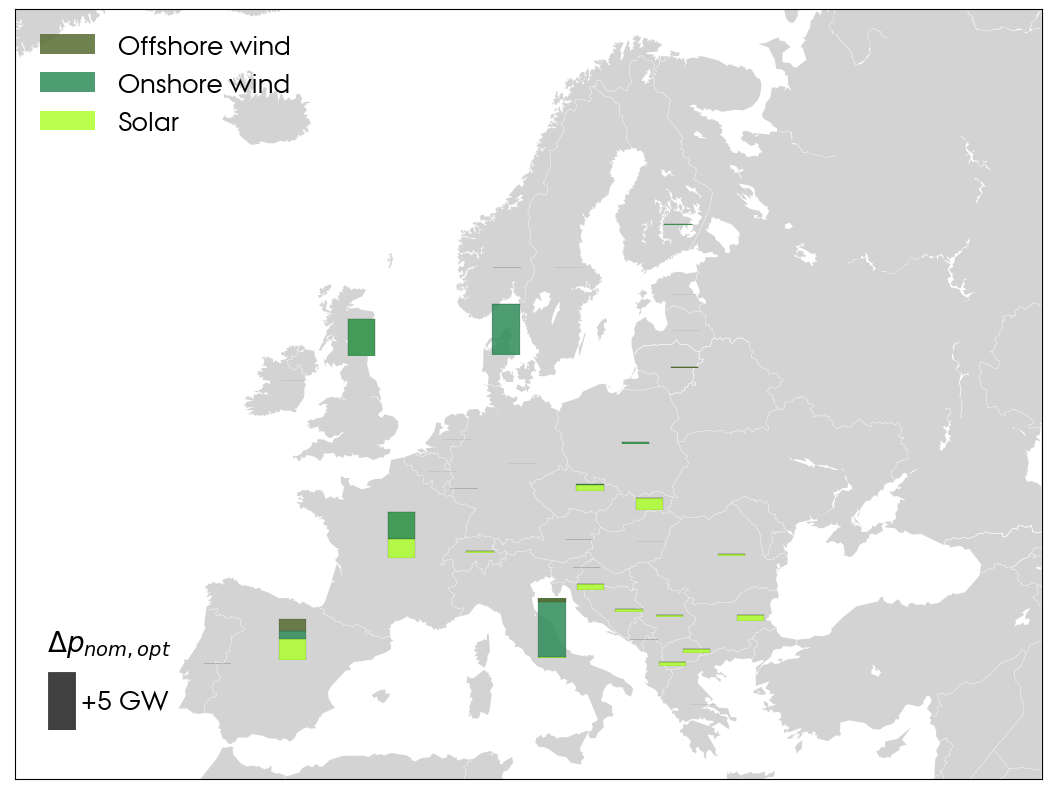}
	\caption{Changes in the optimal wind and solar power capacities due to the climate change effect on hydropower resources.}
	\label{fig:power_capacity_change}
\end{figure}

\vspace{-0.5cm}
\chapter*{References}
\vspace{-1cm}
\printbibliography[heading=none]
\end{refsection}